\PassOptionsToPackage{table,xcdraw}{xcolor}
\documentclass[sigconf]{acmart}

\usepackage{xspace}
\usepackage{natbib}
\usepackage{multirow}
\usepackage[most]{tcolorbox}

\definecolor{green}{RGB}{0,128,0}

\definecolor{yellow}{RGB}{255,200,18}

\newcommand{\stab}{\vspace{1.2ex}\noindent}

\newcommand{\bi}{\begin{itemize}}
\newcommand{\ei}{\end{itemize}}

\newcommand{\be}{\begin{enumerate}}
\newcommand{\ee}{\end{enumerate}}
\newcommand{\beqn}{\begin{eqnarray*}}
\newcommand{\eeqn}{\end{eqnarray*}}

\newcommand{\stitle}[1]{\stab\noindent{\bf #1}}
\newcommand{\etitle}[1]{\vspace{1mm}\noindent{\underline{\textit{#1}}}}

\newcommand{\ie}{\textit{i.e.,} \xspace}
\newcommand{\eg}{\textit{e.g.,} \xspace}

\newcommand{\nlq}{{\sf NL}\xspace}
\newcommand{\sql}{{\sf SQL}\xspace}

\newcommand{\nlsql}{{\sf NL2SQL}\xspace}

\newcommand{\dbs}{{\sf DB}\xspace}

\makeatletter
    \newcommand\figcaption{\def\@captype{figure}\caption}
    \newcommand\tabcaption{\def\@captype{table}\caption}
\makeatother

\NewDocumentCommand{\nan}{ mO{} }{\textcolor{blue}{\textsuperscript{\textit{Nan}}\textsf{\textbf{\small[#1]}}}}

\NewDocumentCommand{\yuyu}{ mO{} }{\textcolor{green}{\textsuperscript{\textit{Yuyu}}\textsf{\textbf{\small[#1]}}}}

\NewDocumentCommand{\boyan}{ mO{} }{\textcolor{purple}{\textsuperscript{\textit{Boyan}}\textsf{\textbf{\small[#1]}}}}

\NewDocumentCommand{\shuyu}{ mO{} }{\textcolor{red}{\textsuperscript{\textit{Shuyu}}\textsf{\textbf{\small[#1]}}}}

\NewDocumentCommand{\xinyu}{ mO{} }{\textcolor{orange}{\textsuperscript{\textit{Xinyu}}\textsf{\textbf{\small[#1]}}}}

\usepackage{marginnote}
\let\oldmarginpar\marginpar
\renewcommand\marginpar[1]{\-\oldmarginpar[\raggedleft\footnotesize #1]%
	{\raggedright\footnotesize\color{blue} #1}} % 注释文字用红色footnote 大小
\usepackage{marginnote}
\let\oldmarginnote\marginnote
\renewcommand\marginnote[1]{\-\oldmarginnote[\raggedleft\footnotesize #1]%
	{\raggedright\footnotesize\color{blue} #1}} %
\newcommand{\dataset}{{\sf {NL2SQL-BUGs}}\xspace}

\newcommand{\spider}{{Spider}\xspace}
\newcommand{\bird}{{BIRD}\xspace}

\usepackage{tcolorbox}

\usepackage{listings}
\usepackage{ulem}
\usepackage{subcaption}
\usepackage{tikz}
\usepackage[edges]{forest}
\usepackage{graphicx} 
\definecolor{hidden-draw}{RGB}{255,255,255}  
\usepackage{longtable}
\newtcolorbox{examplebox}[1]{
 enhanced,
 colback=gray!10,
 colframe=gray!70,
 width=\columnwidth,
 boxrule=0.4pt,
 left=0mm,
 right=0mm,
 top=0mm,
 bottom=0mm,
 before skip balanced=0pt,
 after skip balanced=0pt,
 arc=5pt,
 fonttitle=\bfseries,
 title=#1,
 coltitle=white,  % 使用coltitle替代color
 oversize
}

\AtBeginDocument{%
  }

\copyrightyear{2025}
\acmYear{2025}
\setcopyright{acmlicensed}
\acmConference[KDD '25]{Proceedings of the 31st ACM SIGKDD Conference on Knowledge Discovery and Data Mining V.2}{August 3--7, 2025}{Toronto, ON, Canada}
\acmBooktitle{Proceedings of the 31st ACM SIGKDD Conference on Knowledge Discovery and Data Mining V.2 (KDD '25), August 3--7, 2025, Toronto, ON, Canada}
\begin{document}

\title{NL2SQL-BUGs: A Benchmark for Detecting Semantic Errors in NL2SQL Translation}

\author{Xinyu~Liu}
\orcid{0009-0005-6974-6550}
\affiliation{%
 \institution{The Hong Kong University of Science and Technology (Guangzhou)}
  \city{Guangzhou}
  \country{China}
}
\email{xliu371@connect.hkust-gz.edu.cn}
\author{Shuyu~Shen}
\orcid{0009-0004-7142-6508}
\affiliation{%
  \institution{The Hong Kong University of Science and Technology (Guangzhou)}
  \city{Guangzhou}
  \country{China}
}
\email{sshen190@connect.hkust-gz.edu.cn}

\author{Boyan~Li}
\orcid{0009-0009-8391-4687}
\affiliation{%
  \institution{The Hong Kong University of Science and Technology (Guangzhou)}
  \city{Guangzhou}
  \country{China} 
}
\email{bli303@connect.hkust-gz.edu.cn}

\author{Nan~Tang}
\orcid{0000-0003-2832-0295}
\authornote{Nan~Tang is the corresponding author.}
\affiliation{%
  \institution{The Hong Kong University of Science and Technology (Guangzhou)}
  \city{Guangzhou}
  \country{China}
}
\email{nantang@hkust-gz.edu.cn}

\author{Yuyu~Luo}
\orcid{0000-0001-9530-3327}
\affiliation{%
  \institution{The Hong Kong University of Science and Technology (Guangzhou)}
  \city{Guangzhou}
  \country{China}
}
\email{yuyuluo@hkust-gz.edu.cn}

% --- 作者简称代码 ---
\renewcommand{\shortauthors}{Xinyu Liu, Shuyu Shen, Boyan Li, Nan Tang, \& Yuyu Luo}

\begin{abstract}
Natural Language to SQL (\ie \nlsql) translation is crucial for democratizing database access, but even state-of-the-art models frequently generate semantically incorrect \sql queries, hindering the widespread adoption of these techniques by database vendors. While existing \nlsql benchmarks primarily focus on correct query translation, we argue that a benchmark dedicated to identifying common errors in \nlsql translations is equally important, as accurately detecting these errors is a prerequisite for any subsequent correction -- whether performed by humans or models.
To address this gap, we propose \dataset, \textit{the first} benchmark dedicated to detecting and categorizing semantic errors in \nlsql translation. \dataset adopts a two-level taxonomy to systematically classify semantic errors, covering 9 main categories and 31 subcategories. The benchmark consists of 2,018 expert-annotated instances, each containing a natural language query, database schema, and \sql query, with detailed error annotations for semantically incorrect queries. 
Through comprehensive experiments, we demonstrate that current large language models exhibit significant limitations in semantic error detection, achieving an average detection accuracy of 75.16\%. Specifically, our method successfully detected \textbf{106} errors (accounting for \textbf{6.91\%}) in BIRD, a widely-used \nlsql dataset, which were previously undetected annotation errors. This highlights the importance of semantic error detection in \nlsql systems. The benchmark is publicly available at \url{https://nl2sql-bugs.github.io/}.
\end{abstract}

\begin{CCSXML}
<ccs2012>
   <concept>
       <concept_id>10002951.10002952.10003197</concept_id>
       <concept_desc>Information systems~Query languages</concept_desc>
       <concept_significance>500</concept_significance>
       </concept>
   <concept>
       <concept_id>10010147.10010178.10010179.10010182</concept_id>
       <concept_desc>Computing methodologies~Natural language generation</concept_desc>
       <concept_significance>500</concept_significance>
       </concept>
 </ccs2012>
\end{CCSXML}

\ccsdesc[500]{Information systems~Query languages}
\ccsdesc[500]{Computing methodologies~Natural language generation}

\keywords{Text-to-SQL, Large Language Model, Interface for Databases}

\maketitle

%!TEX root = ../main.tex
\section{Introduction}
\label{sec:intro}
% \blfootnote{*Nan Tang is the corresponding author.}
\begin{figure}[t!]
	\centering
	\includegraphics[width=\columnwidth]{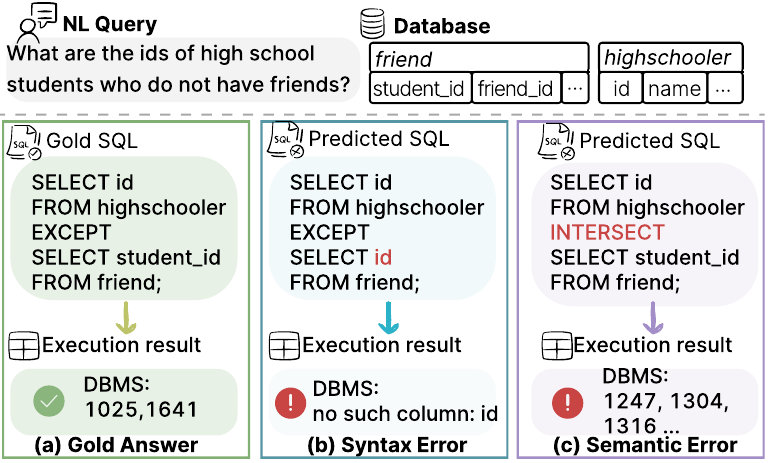}

        \caption{Error Types in Generated NL2SQL Queries.}        
        \Description{This figure shows various error types in NL2SQL query generation.}
	\label{fig:example_errors}
\end{figure}

Over the past few decades, significant progress has been made in translating natural language queries (\nlq) into corresponding \sql queries, commonly referred to as \nlsql (or Text-to-SQL)~\cite{nlsql_survey,nlsql360}, to democratize data analysis~\cite{deepeye_icde,deepeye_sigmod,vis_survey,haichart,DBLP:conf/sigmod/Luo00CLQ21,ncnet,sevi}.
Recent advancements, particularly with large language models (LLMs)~\cite{llmstat,aflow}, have greatly improved the ability to understand complex queries and generate accurate SQL translations~\cite{chung2025longcontextneedleveraging, pourreza2024chasesqlmultipathreasoningpreference, huang2023dataambiguitystrikesback, DBLP:conf/cikm/00090RK24}, leading to better performance on benchmarks like Spider~\cite{dataset-spider} and \bird~\cite{dataset-bird}.
However, despite these advancements, the current state-of-the-art models still achieve only around $75\%$ accuracy on \bird~\cite{dataset-bird}. This indicates that roughly $25\%$ of the cases fail because of \textbf{NL2SQL translation errors}, even in these curated benchmarks. 
In real-world production environments, where databases are more complex~\cite{DBLP:journals/pvldb/ZhangDKKKS23,nlsql360, DBLP:conf/nips/Sivasubramaniam24} and user inputs are more diverse~\cite{fürst2024evaluatingdatamodelrobustness, DBLP:journals/corr/abs-2412-17068}, model performance is likely to degrade further.

\nlsql translation errors can be categorized into two types: \textit{\textbf{syntax errors}} and \textit{\textbf{semantic errors}}. 
Syntax errors, as shown in Figure~\ref{fig:example_errors}(b), occur when the \sql query violates \sql grammar rules or contains invalid references to tables, columns, or operators. These errors are relatively easy to detect because they trigger immediate execution failures and return error messages from the database. 
% For example, Figure~\ref{fig:example_errors}(b) illustrates a case where the generated \sql query attempts to select a non-existent id column from the friend table, resulting in a ``{\tt no such column: id}'' error.

In contrast, \textbf{semantic errors are syntactically correct but fail to reflect the intended meaning of the user's query}, which are thus more subtle and harder to detect.
%\textbf{semantic errors are more subtle and harder to detect.} These errors occur when the \sql query is syntactically valid but fails to reflect the intended meaning of the user's query. 
For example, as shown in Figure~\ref{fig:example_errors}(c), consider a situation where the user asks for students who do not have friends, but the generated \sql query uses {\tt INTERSECT} instead of {\tt EXCEPT}, returning students who have friends rather than those without friends.

%Unlike syntax errors, %which typically cause immediate execution failures, 
%semantic errors lead to incorrect outputs without any visible indication of failure, allowing them to go unnoticed.
All SQL compilers can detect syntax SQL errors, but they cannot detect semantic SQL errors.
As a result, the performance gap in \nlsql systems is largely attributed to semantic errors, which constitute a significant portion of the errors in both the Spider and BIRD datasets. Specifically, after examining the erroneous translations produced by the CodeS \nlsql model~\cite{li2024codes}, we found that 168 out of 170 errors (\textbf{98.8\%}) in Spider and 658 out of 667 errors (\textbf{98.7\%}) in BIRD were semantic. This overwhelming proportion underscores the critical role semantic errors play in \nlsql systems.

\begin{figure}[t!]
\centering
\includegraphics[width=\columnwidth]{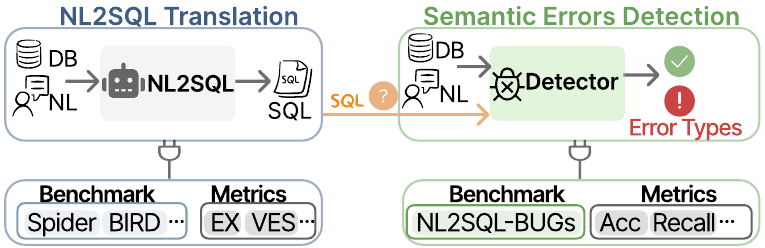}

\caption{Semantic Errors Detection for NL2SQL.}
    \Description{This figure demonstrates semantic errors in NL2SQL query generation.}
\label{fig:Motivation}

\end{figure}

Given the subtle nature of semantic errors, they are more challenging to detect, as they can result in incorrect outputs without any visible errors during execution.
This makes semantic error detection particularly important, as accurately identifying these errors is the \textit{prerequisite} for any subsequent correction. Currently, detecting these errors is often carried out manually by experts. However, to scale this process and reduce human reliance, we need models that can assist or even replace experts in detecting these errors. 
Just as modern DBMSs include \sql syntax checkers to detect and correct syntax errors, \textbf{NL2SQL systems require a similar mechanism -- an NL2SQL semantic detector -- to identify errors in the meaning of generated SQL queries.}  
As shown in Figure~\ref{fig:Motivation}, the task of \textbf{NL2SQL semantic errors detection} aims to detect semantic discrepancies between the natural language query, the generated \sql query, and the database schema. Once semantic errors are detected, these discrepancies can be corrected by triggering error alarms for further inspection, revising the query generation process, or applying automated correction mechanisms.

% To address this gap, we introduce the task of \textbf{NL2SQL semantic errors detection}, as shown in Figure~\ref{fig:Motivation}. Unlike traditional \nlsql translation tasks that focus on generating \sql queries, our task aims to detect semantic discrepancies between the natural language query, the generated \sql query, and the database schema. This approach allows us to identify and categorize various types of semantic errors that may occur during the \nlsql translation process. 
To evaluate the capabilities of \nlsql semantic errors detection, we introduce \dataset, \textit{the first} benchmark designed to identify and classify semantic errors in \nlsql translation.
\dataset adopts a two-level taxonomy to classify semantic errors. It covers nine main categories and numerous subcategories.
\dataset consists of 2,018 expert-annotated instances. Each instance contains a natural language query, a database schema, and the corresponding \sql query, along with detailed error annotations for semantically incorrect \sql queries.

\stitle{Contributions.} We make the following contributions.

\stab {\bf (1) Problem Definition.}  We formally define the task of semantic errors detection in \nlsql translation (Section~\ref{sec: problem_formulation}).

\stab {\bf (2) NL2SQL Semantic Errors Taxonomy.} We introduce a comprehensive two-level taxonomy to classify \nlsql semantic errors into  9 main categories and 31 subcategories (Section~\ref{sec:error}).

\stab {\bf (3) New Benchmark.} We present \dataset, \textit{the first} benchmark designed for evaluating the capabilities of \nlsql semantic errors detection. The dataset consists of 2,018 expert-annotated instances, each with a natural language query, database schema, and \sql query, along with detailed error annotations for semantically incorrect \sql queries (Section~\ref{sec:benchmark}).

\stab {\bf (4) Finding Errors in Existing NL2SQL Benchmarks.} 
We leverage the insights from our error taxonomy and demonstration cases in \dataset to prompt GPT-4o to detect semantic errors in widely-used \nlsql benchmarks -- \spider and \bird. Our evaluation revealed that 16 \sql queries in Spider (\textbf{1.55\%} of the dev set) and \textbf{106} \sql queries in BIRD (\textbf{6.91\%} of the dev set) contained semantic errors that had not been previously identified (Section~\ref{sub:cases}).

\stab{\bf (5) Extensive Experiments.}
We evaluate various LLM-based solutions on the \dataset benchmark to assess their capabilities in detecting \nlsql semantic errors. While some methods perform reasonably well on specific error types (e.g., attribute-related or table-related), overall performance remains limited.
These findings highlight the need for further research into more robust semantic error detection methods (Section~\ref{sec:experiment}).

We contend that \dataset will play an important role in grounding \nlsql techniques for practical applications. We call for cross-community collaboration: database engineers, data mining practitioners, and NLP researchers, to integrate formal logic into training paradigms, fostering ethical, deployable \nlsql systems.
%!TEX root = ../main.tex

\section{Problem and Real-world Cases}

%!TEX root = ../main.tex
\definecolor{colorA}{HTML}{FDDED7} % 浅粉
\definecolor{colorB}{HTML}{E8A6B1} % 浅红粉
\definecolor{colorC}{HTML}{F5BE8F} % 浅橘
\definecolor{colorD}{HTML}{DED5A4} % 淡雅米色
\definecolor{colorE}{HTML}{CCD376} % 浅豆绿
\definecolor{colorF}{HTML}{B0E0D9} % 浅青蓝
\definecolor{colorG}{HTML}{4A9DAF} % 青蓝
\definecolor{colorH}{HTML}{A28CC2} % 浅紫
\definecolor{colorI}{HTML}{8498AB} % 灰蓝

\tikzstyle{my-box}=[
    rectangle,
    draw=hidden-draw,
    rounded corners,
    align=left,
    text opacity=1,
    minimum height=0.5em,
    minimum width=5em,
    fill opacity=.8,
]

\tikzstyle{leaf-head}=[my-box,
    draw=gray,
    fill=gray!10, 
    text=black,
    font=\normalsize,
    inner xsep=1pt,
    inner ysep=1pt,
    line width=0pt,
    align=left,
]

\tikzstyle{leaf-Attribute_Related_Error}=[my-box,
    draw=colorA,
    fill=colorA!25,
    text=black,
    font=\normalsize,
    inner xsep=1pt,
    inner ysep=1pt,
    line width=0pt,
    align=left,
]

\tikzstyle{leaf-Table_Related_Error}=[my-box,
    draw=colorB,
    fill=colorB!25,
    text=black,
    font=\normalsize,
    inner xsep=1pt,
    inner ysep=1pt,
    line width=0pt,
    align=left,
]

\tikzstyle{leaf-Value_Related_Error}=[my-box,
    draw=colorC,
    fill=colorC!25,
    text=black,
    font=\normalsize,
    inner xsep=1pt,
    inner ysep=1pt,
    line width=0pt,
    align=left,
]

\tikzstyle{leaf-Operate_Related_Error}=[my-box,
    draw=colorD,
    fill=colorD!25,
    text=black,
    font=\normalsize,
    inner xsep=1pt,
    inner ysep=1pt,
    line width=0pt,
    align=left,
]
\tikzstyle{leaf-Condition_Related_Error}=[my-box,
    draw=colorE,
    fill=colorE!25,
    text=black,
    font=\normalsize,
    inner xsep=1pt,
    inner ysep=1pt,
    line width=0pt,
    align=left,
]

\tikzstyle{leaf-Function_Related_Error}=[my-box,
    draw=colorF,
    fill=colorF!25,
    text=black,
    font=\normalsize,
    inner xsep=1pt,
    inner ysep=1pt,
    line width=0pt,
    align=left,
]

\tikzstyle{leaf-Clause_Related_Error}=[my-box,
    draw=colorG,
    fill=colorG!25,
    text=black,
    font=\normalsize,
    inner xsep=1pt,
    inner ysep=1pt,
    line width=0pt,
    align=left,
]

\tikzstyle{leaf-Subquery_Related_Error}=[my-box,
    draw=colorH, % 
    fill=colorH!25,
    text=black,
    font=\normalsize,
    inner xsep=1pt,
    inner ysep=1pt,
    line width=0pt,
    align=left,
]

\tikzstyle{leaf-Others}=[my-box,
    draw=colorI,
    fill=colorI!25,
    text=black,
    font=\normalsize,
    inner xsep=1pt,
    inner ysep=1pt,
    line width=0pt,
    align=left,
]
\tikzstyle{lnode-Attribute_Related_Error}=[my-box, minimum height=1.5em,
    draw=colorA,
    fill=white, 
    text=black, font=\normalsize,
    inner xsep=1pt,
    inner ysep=1pt,
    line width=0pt,
]

\tikzstyle{lnode-Table_Related_Error}=[my-box, minimum height=1.5em,
    draw=colorB, 
    fill=white, 
    text=black, font=\normalsize,
    inner xsep=1pt,
    inner ysep=1pt,
    line width=0pt,
]

\tikzstyle{lnode-Value_Related_Error}=[my-box, minimum height=1.5em,
    draw=colorC,
    fill=white, 
    text=black, font=\normalsize,
    inner xsep=1pt,
    inner ysep=1pt,
    line width=0pt,
]
\tikzstyle{lnode-Operate_Related_Error}=[my-box, minimum height=1.5em,
    draw=colorD,
    fill=white, 
    text=black, font=\normalsize,
    inner xsep=1pt,
    inner ysep=1pt,
    line width=0pt,
]

\tikzstyle{lnode-Condition_Related_Error}=[my-box, minimum height=1.5em,
    draw=colorE,
    fill=white, 
    text=black, font=\normalsize,
    inner xsep=1pt,
    inner ysep=1pt,
    line width=0pt,
]

\tikzstyle{lnode-Function_Related_Error}=[my-box, minimum height=1.5em,
    draw=colorF,
    fill=white, 
    text=black, font=\normalsize,
    inner xsep=1pt,
    inner ysep=1pt,
    line width=0pt,
]

\tikzstyle{lnode-Clause_Related_Error}=[my-box, minimum height=1.5em,
    draw=colorG,
    fill=white, 
    text=black, font=\normalsize,
    inner xsep=1pt,
    inner ysep=1pt,
    line width=0pt,
]

\tikzstyle{lnode-Subquery_Related_Error}=[my-box, minimum height=1.5em,
    draw=colorH,
    fill=white, 
    text=black, font=\normalsize,
    inner xsep=1pt,
    inner ysep=1pt,
    line width=0pt,
]

\tikzstyle{lnode-Others}=[my-box, minimum height=1.5em,
    draw=colorI,
    fill=white, 
    text=black, font=\normalsize,
    inner xsep=1pt,
    inner ysep=1pt,
    line width=0pt,
]

\begin{figure*}
    \centering
    \resizebox{1\textwidth}{!}{
        \begin{forest}
            forked edges,
            for tree={
                grow=east,
                reversed=true,
                anchor=base west,
                parent anchor=east,
                child anchor=west,
                base=left,
                font=\normalsize,
                rectangle,
                draw=hidden-draw,
                rounded corners,
                align=left,
                minimum width=1em,
                edge+={darkgray!40, line width=1pt},
                s sep=3pt,
                inner xsep=0pt,
                inner ysep=3pt,
                line width=0pt,
                ver/.style={rotate=90, child anchor=north, parent anchor=south, anchor=center},
            }, 
            [The Taxonomy of\\NL2SQL Translation\\Semantic Errors, leaf-head, text width=8.5em
                [Attribute-related Errors \\ (\S\ref{subsub:Attribute Related Error}), leaf-Attribute_Related_Error, text width=10.5em
                    [Attribute Mismatch, leaf-Attribute_Related_Error, text width=13em
                        [The attribute {[}A{]} may be wrong., lnode-Attribute_Related_Error, text width=33em]
                    ]
                    [Attribute Redundancy, leaf-Attribute_Related_Error, text width=13em
                        [The attribute {[}A{]} may not be mentioned in the \nlq., lnode-Attribute_Related_Error, text width=33em]
                    ]
                    [Attribute Missing, leaf-Attribute_Related_Error, text width=13em
                        [The attribute {[}A{]} may be missing., lnode-Attribute_Related_Error, text width=33em]
                    ]
                ]
            [Table-related Errors \\ (\S\ref{subsub:Table Related Error}), leaf-Table_Related_Error, text width=10.5em
                [Table Mismatch, leaf-Table_Related_Error, text width=13em
                    [The table {[}T{]} may be wrong., lnode-Table_Related_Error, text width=33em]
                ]
                [Table Redundancy, leaf-Table_Related_Error, text width=13em
                    [The table {[}T{]} may be unnecessary., lnode-Table_Related_Error, text width=33em]
                ]
                [Table Missing, leaf-Table_Related_Error, text width=13em
                    [The table {[}T{]} may be missing., lnode-Table_Related_Error, text width=33em]
                ]
                [Join Condition Mismatch, leaf-Table_Related_Error, text width=13em
                    [The join condition between table {[}T{]} and table {[}T{]} is incorrect., lnode-Table_Related_Error, text width=33em]
                ]
                [Join Type Mismatch, leaf-Table_Related_Error, text width=13em
                    [The join type {[}K{]} (e.g.{,} LEFT JOIN) is inconsistent with the \nlq., lnode-Table_Related_Error, text width=33em]
                ]
            ]
            [Value-related Errors\\ (\S\ref{subsub:Value Related Error}), leaf-Value_Related_Error, text width=10.5em
                [Value Mismatch, leaf-Value_Related_Error, text width=13em
                    [The value {[}V{]} in condition {[}C{]} may be wrong., lnode-Value_Related_Error, text width=33em]
                ]
                [Data Format Mismatch, leaf-Value_Related_Error, text width=13em
                    [The data format of value {[}V{]} in attribute {[}A{]} may be wrong., lnode-Value_Related_Error, text width=33em]
                ]
            ]
            [Operator-related Errors\\ (\S\ref{subsub:Operate Related Error}), leaf-Operate_Related_Error, text width=10.5em
                [Comparison Operator, leaf-Operate_Related_Error, text width=13em
                    [The comparison operator {[}O{]} in condition {[}C{]} may be wrong., lnode-Operate_Related_Error, text width=33em]
                ]
                [Logical Operator, leaf-Operate_Related_Error, text width=13em
                    [The boolean operator {[}O{]} or the logical operator precedence may be wrong., lnode-Operate_Related_Error, text width=33em]
                ]
            ]
            [Condition-related Errors\\ (\S\ref{subsub:Condition Related Error}), leaf-Condition_Related_Error, text width=10.5em
                [Explicit Condition  Missing, leaf-Condition_Related_Error, text width=13em
                    [The condition {[}C{]} in \nlq may be missing., lnode-Condition_Related_Error, text width=33em]
                ]
                [Explicit Condition Mismatch, leaf-Condition_Related_Error, text width=13em
                    [The condition {[}C{]} may be wrong., lnode-Condition_Related_Error, text width=33em]
                ]
                [Explicit Condition Redundancy, leaf-Condition_Related_Error, text width=13em
                    [The condition {[}C{]} which not mentioned in \nlq., lnode-Condition_Related_Error, text width=33em]
                ]
                [Implicit Condition Missing, leaf-Condition_Related_Error, text width=13em
                    [The \sql fails to include implicit conditions {[}C{]} (e.g.{,} IS NOT NULL)., lnode-Condition_Related_Error, text width=33em]
                ]
            ]
            [Function-related Errors\\ (\S\ref{subsub:Function Related Error}), leaf-Function_Related_Error, text width=10.5em
                [Aggregate Functions, leaf-Function_Related_Error, text width=13em
                    [The usage of aggregate functions {[}F{]} (e.g.{,} SUM{,} AVG) is incorrect., lnode-Function_Related_Error, text width=33em]
                ]
                [Window Functions, leaf-Function_Related_Error, text width=13em
                    [The usage of window functions {[}F{]} (e.g.{,} OVER{,} PARTITION BY) is incorrect., lnode-Function_Related_Error, text width=33em]
                ]
                [Date/Time Functions, leaf-Function_Related_Error, text width=13em
                    [The usage of date/time functions {[}F{]} (e.g.{,} JULIANDAY{,} strftime) is incorrect., lnode-Function_Related_Error, text width=33em]
                ]
                [Conversion Functions, leaf-Function_Related_Error, text width=13em
                    [The usage of conversion functions {[}F{]} (e.g.{,} CAST) is incorrect., lnode-Function_Related_Error, text width=33em]
                ]
                [Math Functions, leaf-Function_Related_Error, text width=13em
                    [The usage of math functions {[}F{]} (e.g.{,} ROUND) is incorrect., lnode-Function_Related_Error, text width=33em]
                ]
                [String Functions, leaf-Function_Related_Error, text width=13em
                    [The usage of string functions {[}F{]} (e.g.{,} SUBSTR) is incorrect., lnode-Function_Related_Error, text width=33em]
                ]
                [Conditional Functions, leaf-Function_Related_Error, text width=13em
                    [The usage of conditional functions {[}F{]} (e.g.{,} IIF{,} CASE WHEN) is incorrect., lnode-Function_Related_Error, text width=33em]
                ]
            ]
            [Clause-related Errors\\ (\S\ref{subsub:Clause Related Error}), leaf-Clause_Related_Error, text width=10.5em
                [Clause Missing, leaf-Clause_Related_Error, text width=13em
                    [The clause {[}K{]} (e.g.{,} GROUP BY) is missing., lnode-Clause_Related_Error, text width=33em]
                ]
                [Clause Redundancy, leaf-Clause_Related_Error, text width=13em
                    [The clause {[}K{]} (e.g.{,} GROUP BY) is redundancy., lnode-Clause_Related_Error, text width=33em]
                ]
            ]
            [Subquery-related Errors\\ (\S\ref{subsub:Subquery Related Error}), leaf-Subquery_Related_Error, text width=10.5em
                [Subquery Missing, leaf-Subquery_Related_Error, text width=13em
                    [The subquery {[}Q{]} is missing., lnode-Subquery_Related_Error, text width=33em]
                ]
                [Subquery Mismatch, leaf-Subquery_Related_Error, text width=13em
                    [The subquery {[}Q{]} is mismatch with the logic with \nlq., lnode-Subquery_Related_Error, text width=33em]
                ]
                [Partial Query, leaf-Subquery_Related_Error, text width=13em
                    [The query {[}Q{]} is a partial query that contributes to the complete \sql., lnode-Subquery_Related_Error, text width=33em]
                ]
            ]
            [Other Errors\\ (\S\ref{subsub:Other}), leaf-Others, text width=10.5em
                [ASC/DESC, leaf-Others, text width=13em
                    [The usage of ASC/DESC is incorrect., lnode-Others, text width=33em]
                ]
                [DISTINCT, leaf-Others, text width=13em
                    [The usage of DISTINCT is either omitted or incorrectly applied., lnode-Others, text width=33em]
                ]
                [Other, leaf-Others, text width=13em
                    [The \sql generated by the model almost necessitates a complete rewrite., lnode-Others, text width=33em]
                ]
            ]
            ]
            \end{forest}}
    \vspace{-1em}
    \caption{A Taxonomy of NL2SQL Translation Semantic Errors.}
    \label{fig:taxonomy}
    \Description{} 
    % \vspace{-1em}
\end{figure*}
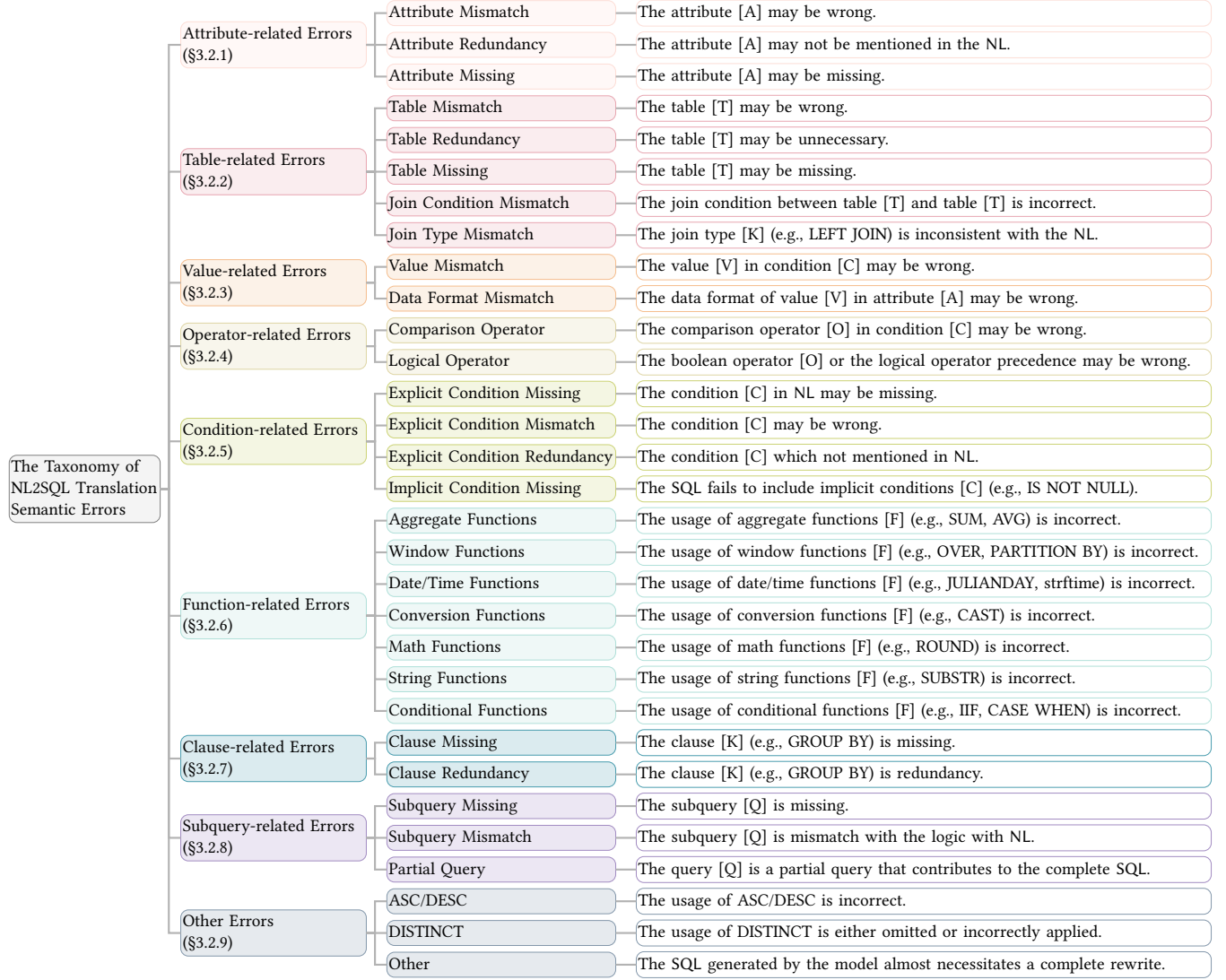

\subsection{Problem Formulation}
\label{sec: problem_formulation}

\stitle{NL2SQL Semantic Error Detection.}
Given a triple (\nlq, \dbs, \sql) consisting of a natural language question (\nlq), a relational database (\dbs), and an SQL query (\sql) generated by models or provided by humans, the task of semantic error detection aims to verify whether \sql is semantically equivalent to the \nlq over \dbs. 
There are two types of semantic errors:
(1) \textit{Incorrect results}: The query returns results that do not match the expected intent of the natural language query.
(2) \textit{Correct results on a specific instance}: The query may produce correct results, but only for the current database instance; the results may not hold for all possible database instances.

Formally, the task can be defined as a function 
$\mathcal{F}$ that maps the triple to a binary decision:
$\mathcal{F}(\text{\nlq}, \text{\dbs}, \text{\sql}) \rightarrow \{True, Fasle\}$, where $True$ (resp. $False$) indicates that $\sql$ is semantically correct (resp. wrong) w.r.t. the given question $\nlq$.

If a semantic error is detected, \ie $\mathcal{F}(\nlq, \dbs, \sql) = False$, the next task is to classify the type of semantic error. This can be formally defined as follows:

\stitle{NL2SQL Semantic Error Type Detection.}
Given a triple (\nlq, \dbs, \sql) consisting of a natural language question (\nlq), a relational database (\dbs), and a generated SQL query (\sql), semantic error classification aims to identify the specific category of semantic error when \sql fails to capture \nlq's intent over \dbs:
$\mathcal{F}(\text{\nlq}, \text{\dbs}, \text{\sql}) \rightarrow T$,
where $T$ is the predefined set of semantic error types, 
$\mathcal{F}(\nlq, \dbs, \sql) = t \in T$ represents the specific type of semantic error identified in the \sql query.

\subsection{Detected Errors in Popular Benchmarks}
\label{sub:cases}

To demonstrate the utility of the semantic error detection task, we applied GPT-4o~\cite{hurst2024gpt} to detect semantic errors in two widely used benchmarks: Spider~\cite{dataset-spider} and BIRD~\cite{dataset-bird}. For each dataset, we used natural language queries (\nlq), database schemas (\dbs), and corresponding SQL queries (\sql) as input to GPT-4o~\cite{hurst2024gpt}. The model then determined whether each \sql query was semantically correct with respect to the natural language query and the database schema.

After the automated semantic error detection, we manually validated the LLM's predictions to ensure accuracy. Our validation revealed that 16 \sql queries in Spider (1.55\% of the development set) and \textbf{106} \sql queries in BIRD (\textbf{6.91\%} of the development set) \textit{contained semantic errors} that had not been previously identified. The real errors identified in both the \spider and \bird datasets are provided in Appendix~\ref{append:errors}.

These findings demonstrate the utility of the semantic error detection model and highlight the importance of devising semantic error detection models to find hidden errors in \nlsql datasets.

%!TEX root = ../main.tex
\section{NL2SQL Semantic Errors Taxonomy}
\label{sec:error}

\subsection{Overview}
The task of detecting semantic errors in \nlsql translation requires a clear and structured understanding of the types of errors that can arise. Based on the extensive analysis of \nlsql systems~\cite{li2023resdsql, li2024codes, talaei2024chess, pourreza2024din, sun2023sql}, we propose a comprehensive two-level taxonomy to categorize semantic errors in \nlsql translation. 

The classification of semantic errors in \nlsql is based on the structure of SQL queries, common translation mistakes, and their impact on query semantics. This approach allows for systematic error identification at various stages of query generation, helping to pinpoint where and why translation mistakes occur. 
Therefore, as shown in Figure~\ref{fig:taxonomy}, we classify semantic errors into 9 main categories such as {\tt Attribute-related Errors}. Each category contains multiple subcategories that identify specific types of errors, allowing for a detailed understanding of how and where translation mistakes occur. 
For example, errors in mapping attributes (columns) from the natural language query to the corresponding attributes in the database are classified under {\tt Attribute-related Errors}.

\subsection{Two-Level Taxonomy}
\label{sub:Two-Level Taxonomy Details}

%This section elaborates on the specifics of the two-level taxonomy, explaining the classification of each error type.

\subsubsection{Attribute-related Errors}
\label{subsub:Attribute Related Error}
Attribute-related Errors occur when NL2SQL models incorrectly map the required columns in natural language queries to corresponding attributes in the database schema. These errors manifest in three primary forms:

\noindent\textbf{Attribute Mismatch}: The model selects incorrect attributes from the schema, indicating a misunderstanding between natural language expressions and their corresponding database fields.

\noindent\textbf{Attribute Redundancy}: The model includes unnecessary attributes not mentioned or implied in the natural language query, suggesting over-interpretation of the input.

\noindent\textbf{Attribute Missing}: The model fails to identify and include attributes that are essential to fulfilling the query intent, indicating an incomplete understanding of the natural language requirements.

These errors primarily stem from the semantic gap between natural language expressions and database schema representations, where models struggle to establish accurate mappings between user intent and the formal database structure.

\subsubsection{Table-related Errors}
\label{subsub:Table Related Error}
Table-related Errors occur when \nlsql models fail to correctly identify, link, or utilize the required tables from the database schema. These errors encompass both basic table selection issues and more complex {\tt JOIN} operations, manifesting in five distinct forms:

\noindent\textbf{Table Mismatch}: The model selects incorrect tables from the schema, indicating a misalignment between natural language query intent and table identification.

\noindent\textbf{Table Redundancy}: The model includes unnecessary tables not required by the query intent, which occurs particularly frequently and suggests over-complication of the query structure.

\noindent\textbf{Table Missing}: The model fails to include tables that are required, leading to an incomplete understanding of the schema.

\noindent\textbf{Join Condition Mismatch}: The model exhibits errors in constructing {\tt JOIN} conditions between tables, creating invalid {\tt JOIN} between unrelated attributes, which demonstrates difficulties in understanding the relationships between tables in the database.

\noindent\textbf{Join Type Mismatch}: The model selects inappropriate {\tt JOIN} types (e.g., {\tt LEFT JOIN}, {\tt INNER JOIN}) that do not align with the natural language query intent, showing a misinterpretation of the desired data inclusion/exclusion patterns.

\begin{examplebox}{Table-Related Example}
\small
\textbf{Query:} List all students and their course grades, including students who haven't taken any courses.

\begin{minipage}[t]{0.45\textwidth}
\textbf{Incorrect SQL:}

SELECT s.name, e.grade

FROM student s

\textcolor{red}{INNER JOIN} enrollment e

\texttt{ON \textcolor{red}{s.id = e.id}}
\end{minipage}
\hfill
\begin{minipage}[t]{0.5\textwidth}
\textbf{Correct SQL:}

SELECT s.name, e.grade

FROM student s

\textcolor{green}{LEFT JOIN} enrollment e

\texttt{ON \textcolor{green}{s.id = e.student\_id}}
\end{minipage}
\end{examplebox}

The example demonstrates both {\tt JOIN} Type Mismatch and {\tt JOIN} Condition Mismatch errors: Using {\tt INNER JOIN} excludes students without courses, contrary to the requirement. The connection condition between the student table and the enrollment table is wrong (\ie \texttt{student.id} $\rightarrow$ \texttt{enrollment.student\_id}).

\subsubsection{Value-related Errors}
\label{subsub:Value Related Error}
Value-related Errors arise when \nlsql models incorrectly parse or interpret the required attribute values from natural language expressions. These errors manifest in two primary forms:

\noindent\textbf{Value Mismatch}: The model fails to map values from natural language descriptions to their actual representations in the database, occurring frequently and reflecting a misunderstanding of how values are actually stored in the database.

\noindent\textbf{Data Format Mismatch}: The model fails to properly format values according to the attribute's data type requirements, showing difficulties in handling type-specific representations.

Please refer to Appendix~\ref{append:value} for an example.

\subsubsection{Operator-related Errors}
\label{subsub:Operate Related Error}
Operator-related Errors occur when \nlsql models fail to select appropriate operators for SQL conditions. These errors manifest in two primary forms:

\noindent\textbf{Comparison Operator}: The model selects incorrect comparison operators in query conditions, such as using `{\tt >}' when `{\tt >=}' is required or confusing `{\tt LIKE}' with `{\tt =}', indicating a semantic gap between natural language intent and SQL comparison operations.

\noindent\textbf{Logical Operator}: The model misinterprets logical relationships expressed in natural language, resulting in the incorrect use of boolean operators (e.g., {\tt AND}, {\tt OR}, {\tt NOT}) in SQL conditions or the logical operator precedence may be wrong.

% Please refer to Section~\ref{append:value} in the Appendix for an example.

\begin{examplebox}{Operator-Related Example}
\small
\textbf{Query:} Find all courses that started after January 1st, 2023 and have more than 30 students.

\begin{minipage}[t]{0.48\textwidth}
\textbf{Incorrect SQL:}

SELECT Student\_ID

FROM Students

WHERE Grade = 3 AND 

Math\_Score > 90 OR 

English\_Score > 90
\end{minipage}
\hfill
\begin{minipage}[t]{0.48\textwidth}
\textbf{Correct SQL:}

SELECT Student\_ID

FROM Students

WHERE Grade = 3 AND 

\textcolor{green}{(} Math\_Score > 90 OR 

English\_Score > 90 \textcolor{green}{)}

\end{minipage}
\end{examplebox}

This example demonstrates Logical Operator errors: By default, the {\tt AND} operator has higher precedence than the {\tt OR} operator. So, the incorrect query is interpreted by the database as: {\tt WHERE (Grade = 3 AND Math\_Score > 90) OR English\_Score > 90}.

% 复合错误
\subsubsection{Condition-related Errors}
\label{subsub:Condition Related Error}
Condition-related Errors occur when \nlsql models fail to properly handle query conditions, particularly in cases where multiple condition-related issues co-exist. These errors are categorized into explicit and implicit conditions:

\noindent\textbf{Explicit Condition Missing}: The model fails to include conditions that are explicitly stated in the natural language query, indicating an incomplete interpretation of the user requirements.

\noindent\textbf{Explicit Condition Mismatch}: The model generates incorrect conditions that deviate from the natural language requirements, showing a misunderstanding of the query intent. A condition typically consists of three parts (attribute, operator, and value), and we classify it as a condition error only when errors occur in two or more components.

\noindent\textbf{Explicit Condition Redundancy}: The model adds unnecessary conditions not mentioned in the natural language query, suggesting an over-interpretation of the query requirements.

\noindent\textbf{Implicit Condition Missing}: The model fails to include necessary conditions that are implied but not explicitly stated in the natural language query, such as {\tt NULL} value handling.

Please refer to Appendix~\ref{append:condition} for an example.

% \input{./example/condition_error_example.tex}

% This example demonstrates explicit and implicit condition errors. The explicit condition mismatch appears in using the wrong attribute (dept) and wrong value (CS) instead of matching the course name, while the implicit condition error manifests in missing the grade {\tt IS NOT NULL} check for completed courses.

\subsubsection{Function-related Errors}
\label{subsub:Function Related Error}
Function-related Errors occur when \nlsql models fail to correctly use SQL functions or misunderstand their purposes. We consider the following cases.

\noindent\textbf{Aggregate Functions}: The model incorrectly applies aggregate functions (\eg {\tt SUM}), such as using them without proper {\tt GROUP BY} clauses or misunderstanding their effect on result sets.

\noindent\textbf{Window Functions}: The model misuses window functions (\eg {\tt OVER}, {\tt PARTITION BY}), often failing to properly specify window frames or partition criteria.

\noindent\textbf{Date/Time Functions}: The model incorrectly handles date and time manipulations (\eg {\tt STRFTIME}), such as wrong format strings or improper date arithmetic.

\noindent\textbf{Conversion Functions}: The model fails to properly convert between data types (\eg {\tt CAST}, {\tt CONVERT}), leading to type mismatch errors or incorrect data transformations.

\noindent\textbf{Math Functions}: The model misapplies mathematical functions (\eg {\tt ROUND}, {\tt CEIL}, {\tt FLOOR}), such as incorrect precision specifications or inappropriate numerical operations.

\noindent\textbf{String Functions}: The model incorrectly uses string manipulation functions (\eg {\tt SUBSTR}, {\tt CONCAT}), such as wrong substring positions or improper string concatenations.

\noindent\textbf{Conditional Functions}: The model misuses conditional functions (\eg {\tt IIF}, {\tt CASE WHEN}), such as incorrect condition logic or improper result expressions.

Please refer to Appendix~\ref{append:function} for an example.

\subsubsection{Clause-related Errors}
\label{subsub:Clause Related Error}

Clause-related Errors occur when \nlsql models fail to correctly include, omit, or structure SQL clauses. These errors primarily involve mishandling {\tt GROUP BY}, {\tt ORDER BY}, and {\tt HAVING} clauses, which are crucial for aggregation, sorting, and filtering grouped results. The errors manifest in two main categories:

\noindent\textbf{Clause Missing}: The model fails to include necessary clauses such as {\tt GROUP BY}, {\tt ORDER BY}, or {\tt HAVING}, resulting in incomplete or incorrect SQL queries that do not fully capture the intent of the natural language.

\noindent\textbf{Clause Redundancy}: The model includes extraneous clauses like {\tt GROUP BY}, {\tt ORDER BY}, or {\tt HAVING} that are not required by the natural language query, indicating an over-interpretation or misunderstanding of the input requirements.

\begin{figure*}[t!]
    \centering    
    
    \includegraphics[width=\linewidth]{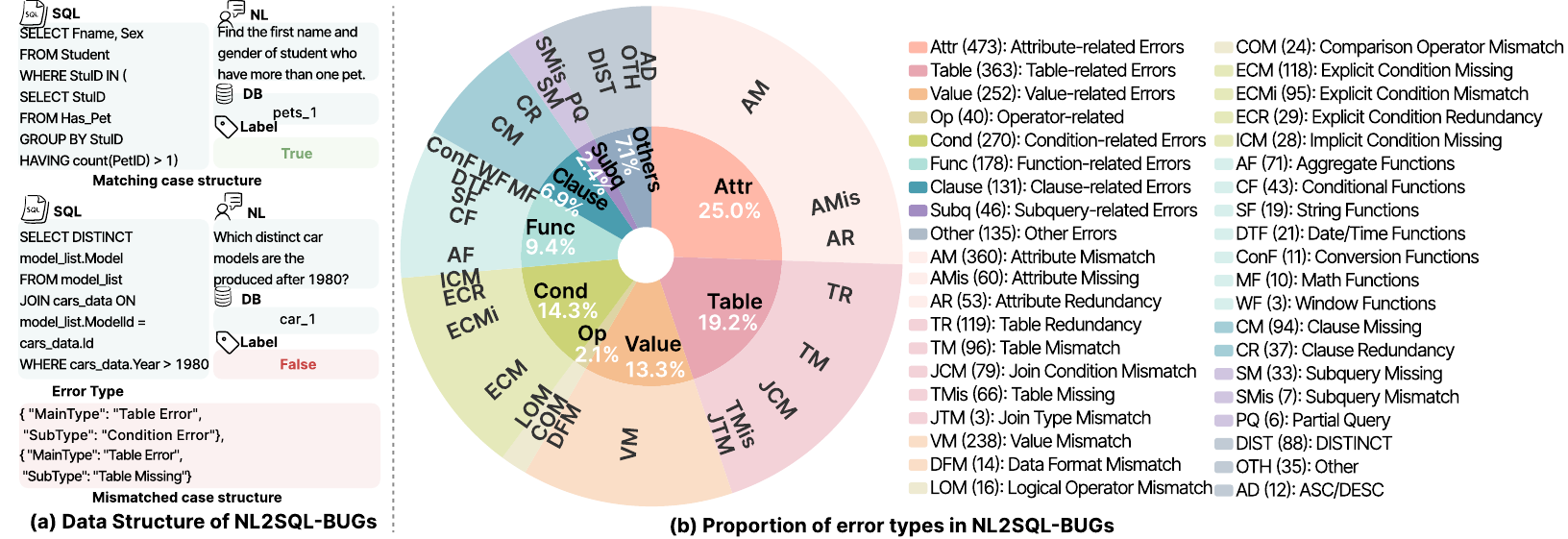}
    \Description{Distribution of different error types in the dataset, showing statistical analysis of the data.}
    
    \caption{Data Structure and Proportion of Error Types in NL2SQL-BUGs Benchmark.}
    \label{fig:error_distribution}
    
\end{figure*}

\subsubsection{Subquery-related Errors}
\label{subsub:Subquery Related Error}
Subquery-related Errors occur when \nlsql models fail to correctly use or generate subqueries. These errors manifest in three distinct categories:

\noindent\textbf{Subquery Missing}: The model fails to generate a necessary subquery, leading to an incomplete or incorrect SQL statement.

\noindent\textbf{Subquery Mismatch}: The model generates a subquery that does not match the context or requirements of the outer query, resulting in logical errors or unexpected results.

\noindent\textbf{Partial Query}: The model generates an SQL query that is only a part of the complete query required to fully address the natural language request. The generated SQL needs to be wrapped inside another query or combined with additional SQL constructs to produce the desired output.
% 
% Please refer to Section~\ref{append:subquery} in the Appendix for an example.

\begin{examplebox}{Subquery-Related Example}
\small
\textbf{Query:} Find the names of students who have enrolled in the maximum number of courses.

\begin{minipage}[t]{0.47\textwidth}
\textbf{Incorrect SQL:}

SELECT COUNT(*) as count, 

student\_id

FROM enrollments

GROUP BY student\_id

ORDER BY count DESC

LIMIT 1
\end{minipage}
\hfill
\begin{minipage}[t]{0.47\textwidth}
\textbf{Correct SQL:}

SELECT s.name FROM students s

JOIN \textcolor{green}{(SELECT COUNT(*) as count,} 
\textcolor{green}{student\_id}

\textcolor{green}{...}

\textcolor{green}{LIMIT 1)}

AS e ON s.id = e.student\_id
\end{minipage}
\end{examplebox}

This example demonstrates a ``Partial Query'' error, where the incorrect SQL query only returns the {\tt student\_id} and course count, but not the students' names. To retrieve the names, an external join with the {\tt students} table is required, making this query a part of a complete query.

\subsubsection{Other Errors}
\label{subsub:Other}
This category is mainly for errors caused by incorrect use of some special keywords in SQL and queries where the SQL intent seriously deviates from the original intent of NL.

\noindent\textbf{ASC/DESC}: This error occurs when the model improperly handles the ordering of query results by either failing to include the correct sorting order ({\tt ASC}/{\tt DESC}) or by incorrectly specifying the order.

\noindent\textbf{DISTINCT}: This error occurs in SQL queries, typically due to the failure to properly apply deduplication operations (e.g., missing {\tt DISTINCT}), resulting in the presence of duplicate records or unnecessary deduplication, which impacts the accuracy and performance of the query results.

\noindent\textbf{Others}: The SQL generated by the model deviates significantly from the required SQL, almost necessitating a complete rewrite. It is difficult to achieve the expected results through corrective modifications. We categorize such errors under this type.

%!TEX root = ../main.tex
\section{NL2SQL-BUGs Overview}
\label{sec:benchmark}

Next, we first overview the characteristics of \dataset (Section~\ref{sub:statis}) and then introduce how \dataset is curated based on real running examples generated by various \nlsql models over popular \nlsql benchmark -- \bird~\cite{dataset-bird} (Section~\ref{sub:construction}).

\subsection{NL2SQL-BUGs Statistics}
\label{sub:statis}

The \dataset benchmark is designed to detect semantic errors in \nlsql translations by distinguishing between correct and incorrect \nlsql translations. 

\stitle{Examples of Errors.}
Figure~\ref{fig:error_distribution}(a) illustrates two examples in \dataset that highlight different types of semantic errors in \nlsql translation. The first example (top-left) represents a correctly generated \sql query, where the natural language query and the SQL query are semantically aligned, resulting in no errors. 
In \dataset, there are a total of \underline{1,019 correct examples}, where the natural language query and the \sql query match semantically and produce correct results.
The second example (bottom-left) highlights a mismatched case, where the \sql query fails to correctly represent the intent of the natural language query, leading to a semantic error. This example is classified as a ``Table Error'' and ``Condition Error'', demonstrating how mismatched table structures or incorrect conditions in the query can produce incorrect results, even when the \sql syntax itself is valid.  
In \dataset, \underline{there are 999 incorrect examples (semantic errors)}, where the \sql queries do not match the intent of the natural language queries, and the discrepancies are categorized into different error types.

\stitle{Error Type Distribution.}
Figure~\ref{fig:error_distribution}(b) shows the distribution of error types within the incorrect translation examples (\ie the \nlsql semantic errors) in the \dataset benchmark. The errors are classified into nine main categories, each capturing a different aspect of semantic mistakes in \nlsql translation. 

The distribution of errors provides several important insights. 
First, attribute-related and table-related errors are the most frequent, highlighting the difficulty models face in correctly identifying and mapping database schema components from natural language queries.
Second, condition-related errors and value-related errors also appear frequently, suggesting that correctly interpreting query conditions and handling values are key areas of challenge.
Third, although function-related and clause-related errors occur less frequently, their impact on \nlsql translation quality is significant, as even a few mistakes in these complex SQL constructs can lead to major semantic discrepancies.

\subsection{NL2SQL-BUGs Construction}
\label{sub:construction}

In this section, we describe how the NL2SQL-Bugs dataset from \bird is curated to support the detection and classification of semantic errors in NL2SQL translations.

\stitle{Dataset Components.} Each instance in \dataset contains the following components:
(1) \textit{Natural Language Query} (\nlq): Expressing various database-related user questions in natural language; 
(2) \textit{Database} (\dbs): The database related to the user question;
(3) \textit{SQL Query} (\sql): Generated translations of \nlq, including both correct and incorrect queries; 
(4) \textit{Correctness Label}: True or False labels indicating whether the SQL query is semantically correct; 
(5) \textit{Error Types}: For incorrect \sql queries, detailed error categorizations are provided to facilitate error detection research.

\stitle{Obtaining Reliable Correctness Labels.}
To establish a robust evaluation benchmark for \nlsql models, obtaining clean correct labels is crucial. The labeling process primarily focuses on execution result matching - whether the SQL query returns identical results as the gold SQL query. For instance in Figure~\ref{fig:error_distribution}(a), consider an SQL query \texttt{``SELECT Fname, Sex FROM Student WHERE StuID IN (SELECT StuID FROM Has\_Pet GROUP BY StuID HAVING COUNT(PetID) > 1)''} paired with the question ``Find the first name and gender of students who have more than one pet.'' This pair is annotated with a True label since its execution results exactly match the gold SQL query. In contrast, for the question ``Which distinct car models are produced after 1980?'' paired with \texttt{``SELECT DISTINCT model\_list.Model FROM model\_list JOIN cars\_data ON model\_list.ModelId = cars\_data.Id WHERE cars\_data.Year > 1980''}, the label would be False as its result set deviates from the gold SQL query due to the incorrect join condition. Through this result-based validation approach, we can establish reliable correctness labels for evaluating NL2SQL models.

\stitle{Determining Error Types.}
To annotate error types for incorrect \sql queries, we employ a two-step approach. 
First, we examine whether the SQL query executes successfully in the database. If it fails to execute, it is classified as a syntax error.
Second, for executable queries that produce incorrect results, we proceed with semantic error classification. Errors are hierarchically classified into main types and subtypes, as detailed in Figure~\ref{fig:taxonomy}. For example, under the main type ``Table-related Error'', we identify subtypes such as ``Join Condition Mismatch'' (when join conditions do not align with the query intent) and ``Table Missing'' (when essential tables are omitted from the query). This hierarchical classification allows for more precise identification of model weaknesses and guides targeted improvements for error correction mechanisms.

\begin{figure}[!t]
	\centering
	\includegraphics[width=\columnwidth]{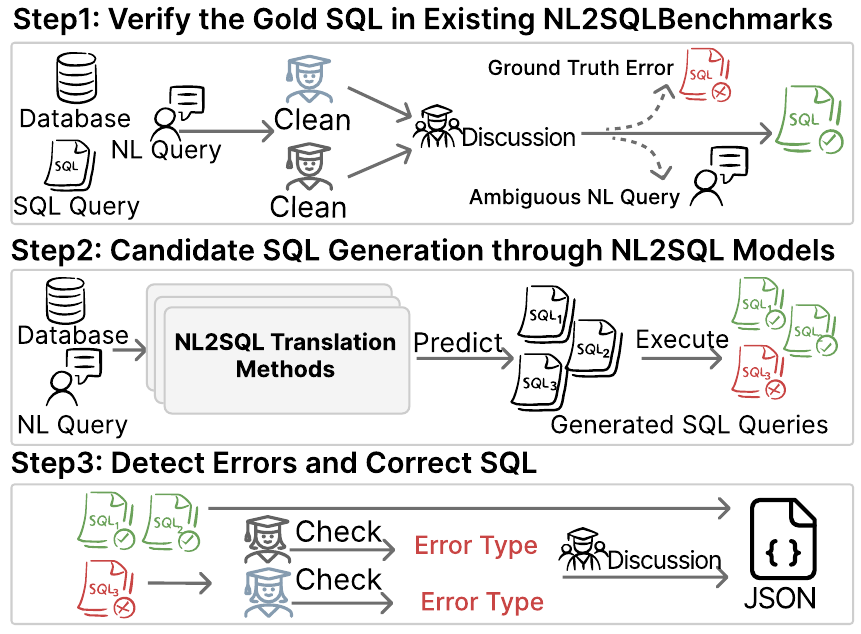}
	\Description{Dataset Construction Workflow.}
	\caption{The Construction Pipeline of NL2SQL-BUGs.}
	\label{fig:Annotation_workflow}
\end{figure}

To ensure SQL correctness and accurate error annotation, we construct \dataset through three steps, as shown in Figure~\ref{fig:Annotation_workflow}.

\etitle{Step 1: Verify the Gold SQL in Existing NL2SQL Benchmarks.}
As discussed in Section~\ref{sub:cases}, existing \nlsql benchmarks may contain semantically incorrect \sql queries, which can compromise the effectiveness of model evaluation. Therefore, the first step in curating \dataset is to verify the ground truth \sql queries from existing benchmarks and correct any errors.

We began by collecting test data from the \bird~\cite{dataset-bird} development dataset, which includes queries of varying difficulty, such as those involving nested queries, aggregate functions, and subqueries. During our inspection, we found that some \sql queries were semantically incorrect, and some \nlq queries lacked a clear corresponding SQL query. To address these issues, we conducted a secondary manual cleaning process.

Specifically, two PhD students were tasked with verifying each SQL query in the dataset, ensuring that they were consistent with both the database schema and the corresponding \nlq. After their individual assessments, they identified errors in \sql queries as well as ambiguous \nlq queries. These cases were then discussed collaboratively until a consensus was reached on the necessary modifications. The finalized erroneous queries were removed, ensuring a high-quality ground truth dataset.

\etitle{Step 2: Candidate SQL Generation through NL2SQL Models.}
After cleaning the dataset, we used multiple \nlsql models from the BIRD~\cite{dataset-bird} leaderboard to generate \sql queries. The models used included RESDSQL~\cite{li2023resdsql} (a pre-trained language model), CodeS~\cite{li2024codes} (a fine-tuned large language model), and CHESS~\cite{talaei2024chess} (an LLM agent with GPT-4o). These models generated \sql queries based on the database schema and the corresponding \nlq query.

Since the first step ensured the correctness of the ground truth data, we directly compared the execution results of the model-generated \sql with the ground truth to determine correctness. In cases where identical \sql queries were generated by multiple models, only one instance was retained to avoid redundant analysis.

\etitle{Step 3: Detect Errors and Correct SQL.}
In the final step, we categorized and annotated the erroneous \sql queries. The annotation labels consisted of 9 major categories and 31 subcategories (see Figure~\ref{fig:taxonomy} for details).
During the annotation process, we provided each database’s ER diagram, requiring the PhD students to first familiarize themselves with it before annotating the \sql errors. Two PhD students independently annotated the errors, and their annotations were aggregated. The Cohen’s Kappa coefficient reached 0.78, indicating a high level of agreement between the annotators. In cases of disagreement about specific error types, discussions were held, and a third \sql expert was consulted to resolve the differences until a consensus was reached.
Both correct samples and incorrect samples with corresponding error types were added to the dataset.

Finally, \dataset contains 1,019 correct examples and 999 incorrect examples with semantic errors.

%!TEX root = ../main.tex

\section{Experiment}
\label{sec:experiment}

In this section, we evaluate the capabilities of \nlsql semantic error detection. Specifically, we try to answer the following questions:
\textbf{Q1:} How effective are different LLMs in detecting \nlsql semantic errors? 
\textbf{Q2:} What are the strengths and weaknesses of various LLMs in detecting different types of semantic errors?

% \stitle{Q3} To what extent can LLMs accurately identify and classify both main error types and subtypes in \nlsql translations?
% \yuyu{discuss Table 2}

%!TEX root = ../main.tex
% \usepackage{multirow}
\begin{table}[t!]
\centering
\caption{Semantic Error Detection of Different LLMs (NP: Negative Precision; NR: Negative Recall; PP: Positive Precision; Positive Recall: PR)}
\label{tab: Error_Detection}
% \usepackage{multirow}
% \vspace{-1em}
\resizebox{\linewidth}{!}{
\begin{tabular}{|c|ccccc|}
\hline
\multirow{2}{*}{\textbf{Models}} &
  \multicolumn{5}{c|}{\textbf{Semantic Error   Detection}} \\ \cline{2-6} 
 &
  \multicolumn{1}{c|}{\textbf{Accuracy}} &
  \multicolumn{1}{c|}{\textbf{NP}} &
  \multicolumn{1}{c|}{\textbf{NR}} &
  \multicolumn{1}{c|}{\textbf{PP}} &
  \textbf{PR} \\ \hline
\textbf{GPT-4o-mini} &
  \multicolumn{1}{c|}{71.56} &
  \multicolumn{1}{c|}{67.43} &
  \multicolumn{1}{c|}{82.28} &
  \multicolumn{1}{c|}{77.85} &
  61.04 \\ \hline
\textbf{GPT-4o} &
  \multicolumn{1}{c|}{\textbf{76.81}} &
  \multicolumn{1}{c|}{72.63} &
  \multicolumn{1}{c|}{85.29} &
  \multicolumn{1}{c|}{82.6} &
  68.5 \\ \hline
\textbf{Claude-3.5-Sonnet} &
  \multicolumn{1}{c|}{76.36} &
  \multicolumn{1}{c|}{71.75} &
  \multicolumn{1}{c|}{86.19} &
  \multicolumn{1}{c|}{83.13} &
  66.73 \\ \hline
\textbf{Gemini-2.0-Flash} &
  \multicolumn{1}{c|}{\textbf{76.81}} &
  \multicolumn{1}{c|}{\textbf{82.66}} &
  \multicolumn{1}{c|}{67.27} &
  \multicolumn{1}{c|}{72.86} &
  \textbf{86.16} \\ \hline
\textbf{DeepSeek-V3} &
  \multicolumn{1}{c|}{75.02} &
  \multicolumn{1}{c|}{70.44} &
  \multicolumn{1}{c|}{85.39} &
  \multicolumn{1}{c|}{81.91} &
  64.87 \\ \hline
\textbf{Qwen2.5-72B} &
  \multicolumn{1}{c|}{74.38} &
  \multicolumn{1}{c|}{69.28} &
  \multicolumn{1}{c|}{\textbf{86.69}} &
  \multicolumn{1}{c|}{\textbf{82.68}} &
  62.32 \\ \hline
\end{tabular}
}
% \vspace{-1em}
\end{table}
% \input{./tables/exp_1_ErrTD.tex}
%!TEX root = ../main.tex
% \usepackage{multirow}
\begin{table*}[t!]
% \vspace{-1em}
\centering
\caption{The Comparison of SQL Error Detection Abilities Across Different Models for Different Error Categories (\(TSA_t\)).}
\label{tab:fineeval}
% \vspace{-1em}
\resizebox{\textwidth}{!}{
\begin{tabular}{|c|c|l|l|l|l|l|l|}
\hline
\textbf{Error Types} &
  \textbf{Sub Error Types} &
  \multicolumn{1}{c|}{\textbf{Qwen2.5-72B}} &
  \multicolumn{1}{c|}{\textbf{GPT-4o}} &
  \multicolumn{1}{c|}{\textbf{GPT-4o-mini}} &
  \multicolumn{1}{c|}{\textbf{Claude-3.5-Sonnet}} &
  \multicolumn{1}{c|}{\textbf{Deepseek-V3}} &
  \multicolumn{1}{c|}{\textbf{Gemini-2.0-Flash}} \\ \hline
 &
  \textbf{Attribute Mismatch} &
  \cellcolor[HTML]{BBE2C7}26.94\% &
  \cellcolor[HTML]{B2DEBF}32.78\% &
  \cellcolor[HTML]{D4ECDC}18.61\% &
  \cellcolor[HTML]{BFE4CA}28.61\% &
  \cellcolor[HTML]{BCE2C8}24.17\% &
  \cellcolor[HTML]{A5D9B4}20.83\% \\ \cline{2-8} 
 &
  \textbf{Attribute Redundancy} &
  \cellcolor[HTML]{EFF7F4}5.66\% &
  \cellcolor[HTML]{DEF0E5}13.21\% &
  \cellcolor[HTML]{E8F4EE}9.43\% &
  \cellcolor[HTML]{ECF6F2}7.55\% &
  \cellcolor[HTML]{F7FAFB}1.89\% &
  \cellcolor[HTML]{FCFCFF}0.00\% \\ \cline{2-8} 
\multirow{-3}{*}{\textbf{Attribute-related   Errors}} &
  \textbf{Attribute Missing} &
  \cellcolor[HTML]{9BD5AB}40.00\% &
  \cellcolor[HTML]{A9DBB7}36.67\% &
  \cellcolor[HTML]{BAE2C7}30.00\% &
  \cellcolor[HTML]{B1DEBF}35.00\% &
  \cellcolor[HTML]{89CE9C}43.33\% &
  \cellcolor[HTML]{85CC99}28.33\% \\ \hline
 &
  \textbf{Table Mismatch} &
  \cellcolor[HTML]{EDF6F2}6.25\% &
  \cellcolor[HTML]{DBEFE3}14.58\% &
  \cellcolor[HTML]{D3ECDC}18.75\% &
  \cellcolor[HTML]{D9EEE1}16.67\% &
  \cellcolor[HTML]{DEF0E5}11.46\% &
  \cellcolor[HTML]{CCE9D6}11.46\% \\ \cline{2-8} 
 &
  \textbf{Table Redundancy} &
  \cellcolor[HTML]{A4D9B3}36.13\% &
  \cellcolor[HTML]{A8DAB7}36.97\% &
  \cellcolor[HTML]{DDF0E4}14.29\% &
  \cellcolor[HTML]{ABDCB9}37.82\% &
  \cellcolor[HTML]{9FD7AF}35.29\% &
  \cellcolor[HTML]{84CC98}28.57\% \\ \cline{2-8} 
 &
  \textbf{Table Missing} &
  \cellcolor[HTML]{83CB96}50.00\% &
  \cellcolor[HTML]{91D1A3}46.97\% &
  \cellcolor[HTML]{CAE8D4}22.73\% &
  \cellcolor[HTML]{A5D9B4}40.91\% &
  \cellcolor[HTML]{7CC890}48.48\% &
  \cellcolor[HTML]{63BE7B}36.36\% \\ \cline{2-8} 
 &
  \textbf{Join Type Mismatch} &
  \cellcolor[HTML]{FCFCFF}0.00\% &
  \cellcolor[HTML]{FCFCFF}0.00\% &
  \cellcolor[HTML]{FCFCFF}0.00\% &
  \cellcolor[HTML]{FCFCFF}0.00\% &
  \cellcolor[HTML]{FCFCFF}0.00\% &
  \cellcolor[HTML]{FCFCFF}0.00\% \\ \cline{2-8} 
\multirow{-5}{*}{\textbf{Table-related   Errors}} &
  \textbf{Join Condition Mismatch} &
  \cellcolor[HTML]{C2E5CD}24.05\% &
  \cellcolor[HTML]{77C78D}58.23\% &
  \cellcolor[HTML]{95D3A7}46.84\% &
  \cellcolor[HTML]{95D3A6}48.10\% &
  \cellcolor[HTML]{80CA94}46.84\% &
  \cellcolor[HTML]{6DC283}34.18\% \\ \hline
 &
  \textbf{Value Mismatch} &
  \cellcolor[HTML]{91D1A3}44.12\% &
  \cellcolor[HTML]{8DCF9F}48.74\% &
  \cellcolor[HTML]{A3D8B2}40.76\% &
  \cellcolor[HTML]{A5D9B4}40.76\% &
  \cellcolor[HTML]{8CCF9E}42.44\% &
  \cellcolor[HTML]{6CC282}34.45\% \\ \cline{2-8} 
\multirow{-2}{*}{\textbf{Value-related   Errors}} &
  \textbf{Data Format Mismatch} &
  \cellcolor[HTML]{DAEEE1}14.29\% &
  \cellcolor[HTML]{CCE9D5}21.43\% &
  \cellcolor[HTML]{DDF0E4}14.29\% &
  \cellcolor[HTML]{CEEAD8}21.43\% &
  \cellcolor[HTML]{D7EDDF}14.29\% &
  \cellcolor[HTML]{DEF0E6}7.14\% \\ \hline
 &
  \textbf{Comparison Operator   Mismatch} &
  \cellcolor[HTML]{E8F4EE}8.33\% &
  \cellcolor[HTML]{C3E5CE}25.00\% &
  \cellcolor[HTML]{EAF5F0}8.33\% &
  \cellcolor[HTML]{C7E7D1}25.00\% &
  \cellcolor[HTML]{FCFCFF}0.00\% &
  \cellcolor[HTML]{EBF5F0}4.17\% \\ \cline{2-8} 
\multirow{-2}{*}{\textbf{Operator-related   Errors}} &
  \textbf{Logical Operator Mismatch} &
  \cellcolor[HTML]{92D1A3}43.75\% &
  \cellcolor[HTML]{99D4A9}43.75\% &
  \cellcolor[HTML]{D3ECDC}18.75\% &
  \cellcolor[HTML]{91D1A3}50.00\% &
  \cellcolor[HTML]{AADBB8}31.25\% &
  \cellcolor[HTML]{93D2A5}25.00\% \\ \hline
 &
  \textbf{Explicit Condition   Missing} &
  \cellcolor[HTML]{63BE7B}62.71\% &
  \cellcolor[HTML]{63BE7B}66.95\% &
  \cellcolor[HTML]{63BE7B}69.49\% &
  \cellcolor[HTML]{63BE7B}71.19\% &
  \cellcolor[HTML]{63BE7B}57.63\% &
  \cellcolor[HTML]{67C07E}35.59\% \\ \cline{2-8} 
 &
  \textbf{Explicit Condition Mismatch} &
  \cellcolor[HTML]{B5DFC1}29.47\% &
  \cellcolor[HTML]{A6D9B5}37.89\% &
  \cellcolor[HTML]{99D4AA}45.26\% &
  \cellcolor[HTML]{B0DDBD}35.79\% &
  \cellcolor[HTML]{A3D8B2}33.68\% &
  \cellcolor[HTML]{97D3A8}24.21\% \\ \cline{2-8} 
 &
  \textbf{Explicit Condition Redundancy} &
  \cellcolor[HTML]{B9E1C5}27.59\% &
  \cellcolor[HTML]{B6E0C2}31.03\% &
  \cellcolor[HTML]{E6F3EC}10.34\% &
  \cellcolor[HTML]{E6F3EC}10.34\% &
  \cellcolor[HTML]{E1F1E8}10.34\% &
  \cellcolor[HTML]{D1EBDA}10.34\% \\ \cline{2-8} 
\multirow{-4}{*}{\textbf{Condition-related   Errors}} &
  \textbf{Implicit Condition Missing} &
  \cellcolor[HTML]{FCFCFF}0.00\% &
  \cellcolor[HTML]{DCEFE3}14.29\% &
  \cellcolor[HTML]{F5F9F9}3.57\% &
  \cellcolor[HTML]{FCFCFF}0.00\% &
  \cellcolor[HTML]{D7EDDF}14.29\% &
  \cellcolor[HTML]{EDF6F3}3.57\% \\ \hline
 &
  \textbf{Aggregate Functions} &
  \cellcolor[HTML]{B8E1C4}28.17\% &
  \cellcolor[HTML]{ACDCBA}35.21\% &
  \cellcolor[HTML]{DAEFE2}15.49\% &
  \cellcolor[HTML]{AEDDBC}36.62\% &
  \cellcolor[HTML]{AADBB9}30.99\% &
  \cellcolor[HTML]{AFDDBD}18.31\% \\ \cline{2-8} 
 &
  \textbf{Window Functions} &
  \cellcolor[HTML]{FCFCFF}0.00\% &
  \cellcolor[HTML]{FCFCFF}0.00\% &
  \cellcolor[HTML]{FCFCFF}0.00\% &
  \cellcolor[HTML]{FCFCFF}0.00\% &
  \cellcolor[HTML]{FCFCFF}0.00\% &
  \cellcolor[HTML]{FCFCFF}0.00\% \\ \cline{2-8} 
 &
  \textbf{Date/Time Functions} &
  \cellcolor[HTML]{DAEEE1}14.29\% &
  \cellcolor[HTML]{E7F4ED}9.52\% &
  \cellcolor[HTML]{F2F8F6}4.76\% &
  \cellcolor[HTML]{DEF0E5}14.29\% &
  \cellcolor[HTML]{B1DEBE}28.57\% &
  \cellcolor[HTML]{D4ECDD}9.52\% \\ \cline{2-8} 
 &
  \textbf{Conversion Functions} &
  \cellcolor[HTML]{FCFCFF}0.00\% &
  \cellcolor[HTML]{FCFCFF}0.00\% &
  \cellcolor[HTML]{FCFCFF}0.00\% &
  \cellcolor[HTML]{FCFCFF}0.00\% &
  \cellcolor[HTML]{E4F3EB}9.09\% &
  \cellcolor[HTML]{FCFCFF}0.00\% \\ \cline{2-8} 
 &
  \textbf{Math Functions} &
  \cellcolor[HTML]{B3DFC0}30.00\% &
  \cellcolor[HTML]{A1D7B1}40.00\% &
  \cellcolor[HTML]{FCFCFF}0.00\% &
  \cellcolor[HTML]{BCE2C8}30.00\% &
  \cellcolor[HTML]{FCFCFF}0.00\% &
  \cellcolor[HTML]{FCFCFF}0.00\% \\ \cline{2-8} 
 &
  \textbf{String Functions} &
  \cellcolor[HTML]{F0F7F4}5.26\% &
  \cellcolor[HTML]{F0F8F5}5.26\% &
  \cellcolor[HTML]{FCFCFF}0.00\% &
  \cellcolor[HTML]{FCFCFF}0.00\% &
  \cellcolor[HTML]{E1F1E7}10.53\% &
  \cellcolor[HTML]{BAE2C6}15.79\% \\ \cline{2-8} 
\multirow{-7}{*}{\textbf{Function-related   Errors}} &
  \textbf{Conditional Functions} &
  \cellcolor[HTML]{EBF6F1}6.98\% &
  \cellcolor[HTML]{FCFCFF}0.00\% &
  \cellcolor[HTML]{FCFCFF}0.00\% &
  \cellcolor[HTML]{F7FAFB}2.33\% &
  \cellcolor[HTML]{F6FAFA}2.33\% &
  \cellcolor[HTML]{FCFCFF}0.00\% \\ \hline
 &
  \textbf{Clause Missing} &
  \cellcolor[HTML]{BCE2C8}26.60\% &
  \cellcolor[HTML]{D1EBDA}19.15\% &
  \cellcolor[HTML]{CBE9D5}22.34\% &
  \cellcolor[HTML]{D6EDDE}18.09\% &
  \cellcolor[HTML]{B9E1C5}25.53\% &
  \cellcolor[HTML]{B5DFC2}17.02\% \\ \cline{2-8} 
\multirow{-2}{*}{\textbf{Clause-related   Errors}} &
  \textbf{Clause Redundancy} &
  \cellcolor[HTML]{C1E4CC}24.32\% &
  \cellcolor[HTML]{EAF5F0}8.11\% &
  \cellcolor[HTML]{DFF0E6}13.51\% &
  \cellcolor[HTML]{F1F8F5}5.41\% &
  \cellcolor[HTML]{D1EBDA}16.22\% &
  \cellcolor[HTML]{C4E5CE}13.51\% \\ \hline
 &
  \textbf{Subquery Missing} &
  \cellcolor[HTML]{E6F4EC}9.09\% &
  \cellcolor[HTML]{E1F1E8}12.12\% &
  \cellcolor[HTML]{FCFCFF}0.00\% &
  \cellcolor[HTML]{F6FAFA}3.03\% &
  \cellcolor[HTML]{ECF6F2}6.06\% &
  \cellcolor[HTML]{D6EDDE}9.09\% \\ \cline{2-8} 
 &
  \textbf{Subquery Mismatch} &
  \cellcolor[HTML]{94D2A5}42.86\% &
  \cellcolor[HTML]{DCEFE3}14.29\% &
  \cellcolor[HTML]{BEE3C9}28.57\% &
  \cellcolor[HTML]{BFE4CB}28.57\% &
  \cellcolor[HTML]{B1DEBE}28.57\% &
  \cellcolor[HTML]{C0E4CC}14.29\% \\ \cline{2-8} 
\multirow{-3}{*}{\textbf{Subquery-related   Errors}} &
  \textbf{Partial Query} &
  \cellcolor[HTML]{FCFCFF}0.00\% &
  \cellcolor[HTML]{FCFCFF}0.00\% &
  \cellcolor[HTML]{FCFCFF}0.00\% &
  \cellcolor[HTML]{FCFCFF}0.00\% &
  \cellcolor[HTML]{FCFCFF}0.00\% &
  \cellcolor[HTML]{FCFCFF}0.00\% \\ \hline
 &
  \textbf{ASC/DESC} &
  \cellcolor[HTML]{6EC385}58.33\% &
  \cellcolor[HTML]{63BE7B}83.33\% &
  \cellcolor[HTML]{63BE7B}58.33\% &
  \cellcolor[HTML]{6DC284}66.67\% &
  \cellcolor[HTML]{63BE7B}58.33\% &
  \cellcolor[HTML]{63BE7B}66.67\% \\ \cline{2-8} 
 &
  \textbf{DISTINCT} &
  \cellcolor[HTML]{FCFCFF}0.00\% &
  \cellcolor[HTML]{EDF6F2}6.82\% &
  \cellcolor[HTML]{FAFBFD}1.14\% &
  \cellcolor[HTML]{F3F9F7}4.55\% &
  \cellcolor[HTML]{FCFCFF}0.00\% &
  \cellcolor[HTML]{FCFCFF}0.00\% \\ \cline{2-8} 
\multirow{-3}{*}{\textbf{Other   Errors}} &
  \textbf{Other} &
  \cellcolor[HTML]{FCFCFF}0.00\% &
  \cellcolor[HTML]{F6FAFA}2.86\% &
  \cellcolor[HTML]{FCFCFF}0.00\% &
  \cellcolor[HTML]{EAF5F0}8.57\% &
  \cellcolor[HTML]{FCFCFF}0.00\% &
  \cellcolor[HTML]{D8EEE0}8.57\% \\ \hline
\end{tabular}
}
\label{table: error_subtype}
% \vspace{-1em}
\end{table*}

\subsection{Experimental Settings}
\label{subsec:expr_setting}

% \stitle{Dataset.}
% \bi
%     \item NL2SQL-BUGs
% \ei

\stitle{Models.}
We consider state-of-the-art LLMs, including open-source models such as Qwen2.5-72B~\cite{qwen2025qwen25technicalreport}, DeepSeek-V3~\cite{liu2024deepseek}, as well as closed-source models like GPT-4o~\cite{hurst2024gpt}, GPT-4o-mini~\cite{o1_mini}, Claude-3.5-Sonnet~\cite{anthropic_claude}, and Gemini-2.0-Flash~\cite{google2025gemini_thinking}. 
For all models, we set the temperature to 0 to ensure deterministic output.

\stitle{Datasets.} We use our curated benchmark \dataset to evaluate LLMs' ability to detect \nlsql semantic errors. It includes 1,019 correct cases and 999 cases with semantic errors.
% All experiments are run on an Ubuntu 22.04.3 LTS server with 512GB of RAM and dual 40-core Intel(R) Xeon(R) Platinum 8383C CPUs (@ 2.70GHz).

\subsection{Experiment for Q1}

The NL2SQL Semantic Error Detection task is formulated as a binary classification task. To evaluate the capability of LLMs in this task, we adopt the classic metrics for classification tasks.

\stitle{Evaluation Metrics.}
We evaluate performance using Overall Accuracy, along with a refined breakdown of Precision and Recall. Precision measures the proportion of correctly predicted True cases (correct SQLs)  out of all cases predicted as True, while Recall assesses the proportion of correctly predicted True cases out of all actual True cases. For the Negative case, the metrics are defined similarly, focusing on correctly identifying incorrect queries.

\stitle{Overall Semantic Error Detection Capability}.
In Table~\ref{tab: Error_Detection}, the performance differences between the models in the NL2SQL semantic error detection task are not highly pronounced. GPT-4o and Claude-3.5-Sonnet are the most balanced models, performing well across various types of semantic errors and outperforming the other models.
Despite achieving a solid overall accuracy of 76.81\%, Gemini-2.0-Flash exhibits imbalanced performance across metrics, as evidenced by its higher positive recall (86.16\%) compared to negative recall (67.27\%), suggesting a greater sensitivity to positives but also a higher risk of false negatives.
DeepSeek-V3 and Qwen2.5-72B still have room for improvement, particularly in enhancing their positive recall and negative precision.

\subsection{Experiment for Q2}
The \nlsql Semantic Error Type Detection focuses on identifying the specific type of error present in an incorrect SQL query. 

\stitle{Evaluation Metrics.}
We report a Type-Specific Accuracy (\(TSA_t\)) for each error type \(t\).
Given a dataset \(D\) of instances and a set \(T\) of all error types, we denote the set of true error types present in an instance \(i \in D\) as \(E_i \subseteq T\), and the set of error types predicted by the model as \(P_i \subseteq T\).  The Type-Specific Accuracy (\(TSA_t\)) for a given error type \(t \in T\) is then defined as:

\[
TSA_t = \frac{|\{i \in D \mid t \in E_i \land t \in P_i\}|}{|\{i \in D \mid t \in E_i\}|}
\]

\begin{figure}[t!]
	\centering
	\includegraphics[width=\columnwidth]{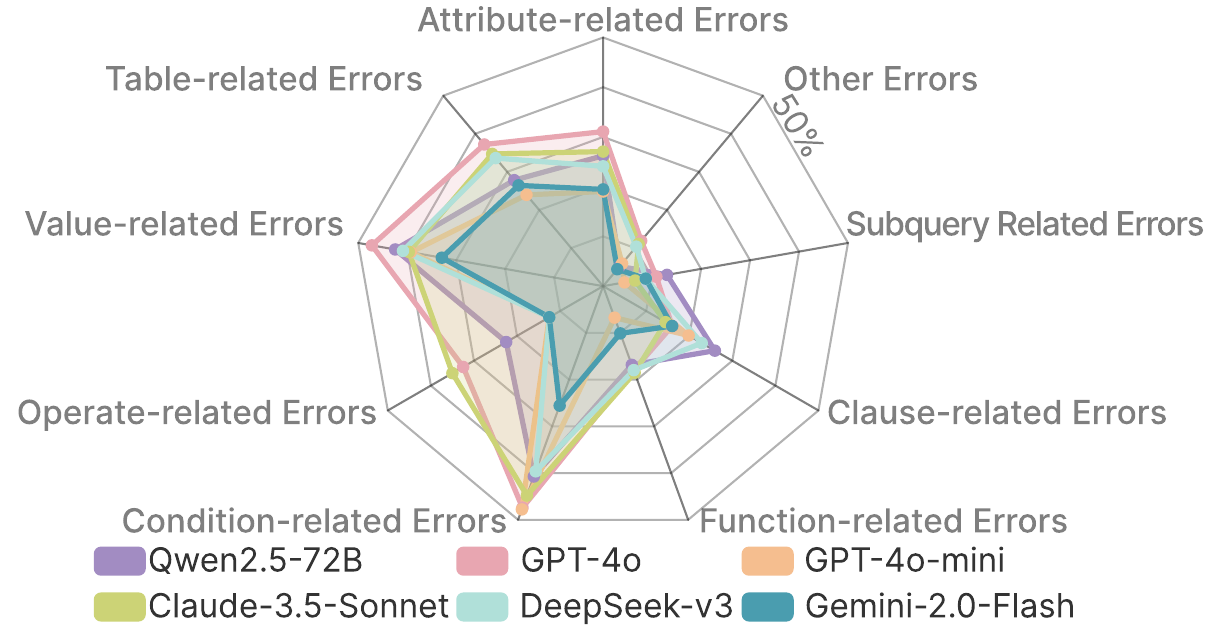}
	\Description{A bar chart comparing NL2SQL models' error detection performance across main error types.}

	\caption{Performance of NL2SQL models in error type detection across main error types (\(TSA_t\)).}
	\label{fig:main_error_types_tsa}
\end{figure}

We calculated the metrics \(TSA_t\) for specific categories, as shown in Table~\ref{table: error_subtype}, which shows the performance of different models across subtypes. The radar chart (see Figure~\ref{fig:main_error_types_tsa}) further demonstrates the performance differences of the models at the main category level.

\stitle{Overall Semantic Type Error Detection Capability}.
Despite achieving relatively high accuracy (around 75\%) in distinguishing correct and incorrect SQL queries(Table~\ref{tab: Error_Detection}), the models fail to reach 50\% accuracy in identifying specific error types (Figure~\ref{fig:main_error_types_tsa}). This discrepancy suggests that while the models may correctly flag a query as erroneous, they often struggle to pinpoint the underlying cause or fail to recognize all existing errors.

The empirical evidence presented in Figure~\ref{fig:main_error_types_tsa} demonstrates that the models exhibit pronounced proficiency in identifying Condition-related errors and Value errors. Conversely, the models display markedly diminished capability in detecting subquery-related errors and those categorized as ``Other Errors'', indicating a significant performance differential across error typologies.

In comparative analysis, GPT-4o and Gemini-2.0-Flash demonstrate identical accuracy metrics (76.81\% in Table~\ref{tab: Error_Detection}); however, regarding error categorization capabilities, GPT-4o exhibits substantially superior performance compared to Gemini-2.0-Flash (Figure~\ref{fig:main_error_types_tsa}). 
This observation illuminates a critical distinction in language model evaluation methodologies: equivalent accuracy metrics do not necessarily reflect equivalent capabilities. GPT-4o demonstrates a deeper understanding of SQL semantics and structure, enabling it not only to detect anomalies (Error Detection), but also to analytically identify the underlying error taxonomies (Error Type Detection).

\stitle{Fine-grained Evaluation of Semantic Error Detection Capability.}
We conducted a fine-grained evaluation of semantic error detection capabilities, with results summarized in Table~\ref{tab:fineeval}. 
The model exhibits deficiencies in fine-grained detection capabilities. For example, Claude-3.5-Sonnet outperforms GPT-4o-mini in many subtypes of error detection, yet underperforms in the detection of Explicit Condition Mismatch, highlighting certain limitations of the model. It demonstrates greater proficiency in handling tasks involving substantial semantic variations. However, its performance is suboptimal, particularly in tasks such as Explicit Condition Missing, which require less complex semantic processing. Moreover, the model struggles with errors that necessitate DB knowledge, such as Join Type Mismatch and Function-Related Errors, revealing considerable potential for improvement.

%!TEX root = ../main.tex
\section{Related Work}
\label{sec:related}

\stitle{NL2SQL Translation Methods.}
\nlsql translation has been an active research area in recent years. Early research relied mainly on rule-based templates and semantic parsing approaches~\cite{10.1145/2588555.2594519, DBLP:conf/sigmod/Katsogiannis-Meimarakis21}, which often struggled with complex queries. With the advancement of deep learning, neural network-based methods~\cite{xiao2016sequence, lin2019grammar, bogin2019representing} substantially improved translation accuracy. The introduction of pre-trained language models made a particular breakthrough~\cite{li2023resdsql}. More recently, Large Language Models (LLMs), which have advanced many fields~\cite{agent_survey,gaivis,spo,datamosaic,verifyai,symphony,zhang-etal-2024-mar}, have made significant strides in \nlsql~\cite{nlsql_survey,nlsql360, elliesql}, pushing accuracy close to 90\% on Spider while achieving 70\% on the more challenging BIRD benchmark~\cite{dataset-bird}. 
While these advancements are impressive, the 30\% error rate in complex scenarios in BIRD highlights significant challenges in real-world deployments, where near-perfect accuracy is crucial. For instance, although models such as Alpha-SQL~\cite{alphsql} and CHASE-SQL~\cite{pourreza2024chasesqlmultipathreasoningpreference} utilize LLMs to achieve notable performance in the NL2SQL domain, they are still prone to significant issues with hallucination and errors.  In critical fields like finance and healthcare, even small query errors can have serious consequences. Despite this, current research mainly focuses on improving overall accuracy, neglecting systematic error analysis and correction.

\stitle{NL2SQL Errors Detection.}
\nlsql faces challenges in understanding user intent, handling complex database schemas, and generating intricate SQL~\cite{nlsql_survey}. Prior approaches to error analysis in \nlsql systems can be broadly categorized into three main types: NL explanation, human-machine collaboration, and visual analytics. DIY~\cite{diy} employs templates to provide step-by-step NL explanations of SQL, helping users to comprehend them incrementally. NaLIR~\cite{nalir}, on the other hand, enables SQL explanation and error detection by creating entity mappings between NL and SQL.
MISP~\cite{mips} adopts a human-machine collaborative approach, leveraging parsed states and incorporating human feedback for SQL error detection.
SQLVis~\cite{sqlvis} applies visualization techniques to SQL error detection by transforming SQL into graphical representations, helping users understand complex SQL and detect errors. 
% 
%However, previous works struggle with reliable semantic error detection and explanations. 
Prior research has faced challenges in reliable semantic error detection and explanation of SQL queries, often relying heavily on manual effort. There has been a preliminary exploration of automated error detection systems~\cite{yang2025SQLDriller}.
To facilitate this research direction, we propose a two-level taxonomy with 9 main categories and 31 subcategories to categorize semantic errors in \nlsql translation. Based on this, we also developed the first benchmark for evaluating \nlsql semantic error detection, containing 2,018 expert-annotated examples.

%!TEX root = ../main.tex
\section{Conclusion}
\label{sec:conclusion}

In this paper, we introduced the task of semantic error detection in \nlsql translation and demonstrated its importance by uncovering 106 errors (6.91\%) in the BIRD benchmark and 16 errors (1.55\%) in the Spider benchmark. We proposed a two-level taxonomy categorizing semantic errors into 9 main categories and 31 subcategories. Based on this taxonomy, we developed \dataset, the first benchmark designed specifically for evaluating \nlsql semantic error detection, containing 2,018 expert-annotated examples.
Through extensive experiments, we highlighted the limitations of current LLM-based models for detecting \nlsql semantic errors. 
This study aims to advance the development of techniques for the automatic detection of semantic errors.

\begin{acks}
This paper is supported by NSF of China (62402409), Guangdong provincial project 2023CX10X008, Guangdong Basic and Applied Basic Research Foundation (2023A1515110
545), Guangzhou Basic and Applied Basic Research Foundation
(2025A04J3935), and Guangzhou-HKUST(GZ) Joint Funding Program (2025A03J3714).
\end{acks}

\bibliographystyle{ACM-Reference-Format}
\balance
\bibliography{main}

% \clearpage

\appendix

%!TEX root = ../main.tex
\section{Detected Errors in NL2SQL Benchmarks}
\label{append:errors}

In Table~\ref{tab: ErrorInBirdAndSPider} and Table~\ref{tab: Error example.}, we list SQL queries with semantic errors found in the BIRD and Spider benchmarks.

\begin{itemize}
    \item Table~\ref{tab: ErrorInBirdAndSPider} lists the indices of SQL queries from the BIRD~\cite{dataset-bird} and Spider~\cite{dataset-spider} development benchmarks that we identified as containing semantic errors. The presence of such errors can cause query results to deviate from the expected answers derived from the natural language question (\nlq).
    \item Table~\ref{tab: Error example.} illustrates specific instances of the semantic errors identified, providing concrete examples. Each example comprises the \nlq, the incorrect SQL statement, and color-coded annotations that highlight the erroneous segments alongside the correct query intent. The showcased errors exemplify various categories, such as redundant fields, incorrect sorting directions, and improper filtering conditions.
\end{itemize}

\begin{table}[!b]
% \small
\centering
\caption{Semantic Errors in BIRD and Spider Benchmarks.}
\label{tab: ErrorInBirdAndSPider}
\renewcommand{\arraystretch}{1.2} %
\setlength{\tabcolsep}{5pt} % 
\begin{tabular}{|c|p{6cm}|} % 
\hline
Benchmark & Index of (\nlq, \sql) Examples with Semantic Errors \\ \hline
\spider & 
  67, 101, 494, 555, 579, 773, 774, 811, 812, 819, 820, 128, 129, 961, 962, 177 \\ \hline
\bird & 
  1027, 1029, 519, 523, 530, 23, 70, 72, 584, 1107, 600, 602, 603, 94, 1119, 1120, 1121, 631, 632, 635, 125, 639, 640, 129, 642, 646, 649, 144, 145, 656, 1170, 667, 679, 682, 1197, 686, 687, 1199, 1204, 693, 182, 186, 194, 1219, 709, 710, 1225, 1233, 1243, 221, 1247, 1248, 1256, 1265, 1269, 247, 1273, 1274, 252, 254, 1279, 1284, 271, 1300, 281, 1308, 296, 1322, 812, 309, 341, 342, 343, 855, 349, 360, 1388, 386, 387, 388, 389, 398, 406, 1450, 1454, 431, 1458, 441, 442, 443, 446, 447, 966, 458, 970, 1482, 973, 978, 1491, 986, 993, 484, 1000, 1004, 1530, 1531 \\ \hline
\end{tabular}
\end{table}
\begin{table*}[!t]
% \small
\centering
\caption{Samples of Detected Semantic Errors in Real-world Benchmarks.}
\label{tab: Error example.}
\resizebox{\textwidth}{!}{
\begin{tabular}{|p{2.5cm}|p{6cm}|p{8cm}|}
\hline
\textbf{Source} & \textbf{NL Query} & \textbf{SQL Query} \\ 
\hline
Spider-494 & List the name, date and \textcolor{green}{result} of each battle. & 
\texttt{%
  \begin{tabular}[t]{@{}l@{}}
  SELECT name , date  \textcolor{red}{t1.age \#Need result} FROM battle;
  \end{tabular}
} \\ 
\hline
Spider-555 & What is the mobile phone number of the student named \textcolor{green}{Timmothy Ward}? & 
\texttt{%
  \begin{tabular}[t]{@{}l@{}}
  SELECT cell\_mobile\_number FROM students \\
  WHERE first\_name = '\textcolor{red}{t}immothy' \textcolor{red}{\#Need 'Timmothy'} \\
    AND last\_name = '\textcolor{red}{w}ard'; '\textcolor{red}{\#Need 'Ward'}
  \end{tabular}
} \\
\hline
Spider-579 & What are the \textcolor{green}{different addresses} that have students living there? & 
\texttt{%
  \begin{tabular}[t]{@{}l@{}}
  SELECT \textcolor{red}{count}(DISTINCT current\_address\_id)\\
  FROM Students;\textcolor{red}{\#No Need count}
  \end{tabular}
} \\
\hline
Spider-773 & What are the countries that have \textcolor{green}{greater surface area} than any country in Europe? & 
\texttt{%
  \begin{tabular}[t]{@{}l@{}}
  SELECT Name FROM country WHERE SurfaceArea > ( \\
    SELECT \textcolor{red}{min}(SurfaceArea) \textcolor{red}{\#max(...)}\\ 
    FROM country WHERE Continent = "Europe"); 
  \end{tabular}
} \\
\hline
BIRD-252 & What are the \textcolor{green}{atoms} that can bond with the atom that has the element lead? \textcolor{gray}{Evidence: Atom that has the element lead refers to atom\-id where element = `pb';} & 
\texttt{%
\begin{tabular}[t]{@{}l@{}}
SELECT T2.atom\-id, \textcolor{red}{T2.atom\_id2 \#No Need} \\
FROM atom AS T1 \\
INNER JOIN connected AS T2 ON T1.atom\-id = T2.atom\-id \\
WHERE T1.element = 'pb';
\end{tabular}
} \\ 
\hline
BIRD-812 & List down at least five \textcolor{green}{full names} of superheroes with blue eyes. \textcolor{gray}{Evidence: Blue eyes refers to colour.colour = 'Blue' WHERE eye\-colour\-id = colour.id;} & 
\texttt{%
  \begin{tabular}[t]{@{}l@{}}
  SELECT \textcolor{red}{T1.superhero\-name \#Need T1.full\_name}\\
  FROM superhero AS T1 \\
  INNER JOIN colour AS T2 ON T1.eye\-colour\-id = T2.id \\
  WHERE T2.colour = 'Blue' LIMIT 5;
  \end{tabular}
} \\ 
\hline
BIRD-986 & In which \textcolor{green}{race} did the fastest 1st lap time was recorded? \textcolor{gray}{Evidence: Please indicate the time in milliseconds. Fastest refers to Min(time);} & 
\texttt{%
  \begin{tabular}[t]{@{}l@{}}
  SELECT \textcolor{red}{T1.milliseconds \#Need races.name}\\
  FROM lapTimes AS T1 \\
  INNER JOIN races AS T2 ON T1.raceId = T2.raceId \\
  WHERE T1.lap = 1 ORDER BY T1.time LIMIT 1;
  \end{tabular}
} \\ 
\hline
BIRD-1029 & What are the speed in which attacks are put together of the \textcolor{green}{top} 4 teams with the highest build Up Play Speed? \textcolor{gray}{Evidence:Speed in which attacks are put together refers to buildUpPlaySpeed;highest build up play speed refers to MAX(buildUpPlaySpeed);} & 
\texttt{%
  \begin{tabular}[t]{@{}l@{}}
  SELECT t1.buildUpPlaySpeed \\
  FROM Team\_Attributes AS t1 INNER JOIN Team AS t2 \\
  ON t1.team\_api\_id = t2.team\_api\_id \\
  ORDER BY t1.buildUpPlayDribbling \textcolor{red}{ASC \#Need DESC} \\
  LIMIT 4;
  \end{tabular}
} \\ 
\hline
BIRD-1248 & How many patients born after 1980/1/1 have an \textcolor{green}{abnormal} fibrinogen level? \textcolor{gray}{Evidence: Born after 1980/1/1 refers to Birthday > '1980-01-01'; \textcolor{green}{normal} fibrinogen level refers to FG \textcolor{green}{between 150 and 450}; Should return the number of distinct patients.} & 
\texttt{%
  \begin{tabular}[t]{@{}l@{}}
  SELECT COUNT(DISTINCT T1.ID) FROM Patient AS T1 \\
  INNER JOIN Laboratory AS T2 ON T1.ID = T2.ID \\
  WHERE T2.FG <\textcolor{red}{=} 150 OR T2.FG >\textcolor{red}{=} 450\\
  \textcolor{red}{\#Not(>=150 and <=450) is <150 or >450}\\
  AND T1.Birthday > '1980-01-01';
  \end{tabular}
} \\ 
\hline
\end{tabular}
}
\end{table*}

\section{Examples in Our Error Taxonomy}
\label{append:exam_tax}

\subsection{Value-related Errors}
\label{append:value}

 This example demonstrates both data format and value mismatch errors. The model incorrectly uses the MM/DD/YY format (`01/01/23') instead of the database's standard YYYY-MM-DD format (`2023-01-01'), while also treating numeric values as strings (`30' instead of 30).
 
% Please refer to Section~\ref{append:value} in the Appendix for an example.

\begin{examplebox}{Value-Related Example}
\small
\textbf{Query:} Find all courses that started after January 1st, 2023 and have more than 30 students.

\begin{minipage}[t]{0.48\textwidth}
\textbf{Incorrect SQL:}

SELECT c.name

FROM course c

WHERE c.start\_date >\textcolor{red}{`01/01/23'}

AND c.enrollment > \textcolor{red}{`30'}
\end{minipage}
\hfill
\begin{minipage}[t]{0.48\textwidth}
\textbf{Correct SQL:}

SELECT  c.name

FROM    course c

WHERE   c.start\_date > \textcolor{green}{`2023-01-01'}
AND c.enrollment > \textcolor{green}{30}
\end{minipage}
\end{examplebox}

\subsection{Condition-related Errors}
\label{append:condition}

This example demonstrates explicit and implicit condition errors. The explicit condition mismatch appears in using the wrong attribute (dept) and wrong value (CS) instead of matching the course name, while the implicit condition error manifests in missing the grade {\tt IS NOT NULL} check for completed courses.

\begin{examplebox}{Condition-Related Example}
\small
\textbf{Query:} Find the students' names and their grades in the Database course.

\begin{minipage}[t]{0.45\textwidth}
\textbf{Incorrect SQL:}

SELECT  s.name, e.grade

FROM student AS s

JOIN enrollment AS e

ON s.id = e.student\_id

JOIN course AS c

ON e.course\_id = c.id

WHERE   \textcolor{red}{c.dept = `CS'}
\end{minipage}
\hfill
\begin{minipage}[t]{0.45\textwidth}
\textbf{Correct SQL:}

SELECT  s.name, e.grade

FROM    student AS s

JOIN enrollment AS e

ON s.id = e.student\_id

JOIN course AS c

ON e.course\_id = c.id

WHERE   \textcolor{green}{c.name = `Database'}

AND \textcolor{green}{e.grade IS NOT NULL}
\end{minipage}
\end{examplebox}

\subsection{Function-related Errors}
\label{append:function}

This example illustrates two function-related errors: using the {\tt AVG} aggregate function in the {\tt WHERE} clause instead of the appropriate {\tt HAVING} clause, and neglecting to use the {\tt ROUND} math function to define decimal precision.

\begin{examplebox}{Function-Related Example}
\small
\textbf{Query:} Calculate the average grade (with two decimal places) for each course, and list courses with averages above 85.

\begin{minipage}[t]{0.47\textwidth}
\textbf{Incorrect SQL:}

SELECT  c.name,

AVG(e.grade)

FROM    course c

JOIN enrollment e

ON c.id = e.course\_id

WHERE  \textcolor{red}{AVG(e.grade) > 85}

GROUP BY c.name
\end{minipage}
\hfill
\begin{minipage}[t]{0.47\textwidth}
\textbf{Correct SQL:}

SELECT  c.name,

\textcolor{green}{ROUND(AVG(e.grade), 2)}

FROM    course c

JOIN enrollment e

ON c.id = e.course\_id

GROUP BY c.name

HAVING  \textcolor{green}{AVG(e.grade) > 85}
\end{minipage}
\end{examplebox}

% \section{Details of Dataset Construction}
% \label{append:build}

% \begin{figure}[htbp]
% 	\centering
% 	\includegraphics[width=\columnwidth]{./figures/Annotation_Workflow_v2.pdf}
% 	\Description{Dataset Construction Workflow.}
% 	\caption{The Construction Pipeline of NL2SQL-BUGs.}
% 	\label{fig:Annotation_workflow}
% \end{figure}

% Figure~\ref{fig:Annotation_workflow} illustrates the dataset construction workflow for NL2SQL-BUGs. The process involves multiple stages, ensuring the quality and diversity of the dataset for NL2SQL tasks. 
% \begin{itemize}
%     \item In \textbf{Step 1}, We begin by collecting test data from the \bird~\cite{dataset-bird} development dataset. Each input natural language (NL) query in the dataset is mapped to its corresponding SQL ground truth. Experts perform data cleaning to ensure correctness, and discussions are conducted to resolve ambiguities, ultimately leading to the final verified SQL query.
%     \item In \textbf{Step 2}, the NL query is processed by an NL2SQL translation model, which predicts and executes SQL queries. The generated queries may be correct (green) or incorrect (red).
%     \item In \textbf{Step 3}, generated SQL queries are checked for errors. If an error is detected, its type is analyzed, followed by discussions to resolve the issue. The corrected SQL query is then converted into JSON format.
% \end{itemize}

% \section{}
% \label{append:errors}

\end{document}